\documentclass[sigconf]{acmart}
\usepackage{mdframed}                  
\usepackage{fontawesome}
\usepackage{enumitem}
\usepackage{booktabs}
\usepackage{diagbox}
\usepackage{colortbl}
\usepackage[table,x11names,dvipsnames]{xcolor} 
\usepackage{subcaption}                        
\usepackage{algorithm}
\usepackage{algpseudocode}
\usepackage{tikz}
\usepackage{pgfplots}
\usepackage{pgfplotstable}
\usepgfplotslibrary{groupplots}
\usepgfplotslibrary{dateplot}
\pgfplotsset{compat=1.18}
\usepackage{multirow, makecell, array, graphicx}
\usepackage{tcolorbox}
\usepackage{lipsum} 
\usetikzlibrary{matrix, positioning, fit}

\definecolor{mygreen}{RGB}{124,205,124}
\definecolor{myred}{RGB}{238,162,173}
\definecolor{myblue}{RGB}{92,172,238}
\definecolor{mygray}{gray}{0.8}
\definecolor{mylightgreen}{RGB}{226, 255, 233}
\definecolor{mydarkgreen}{RGB}{161, 240, 180}
\definecolor{mylightred}{RGB}{255, 232, 230}
\definecolor{mydarkred}{RGB}{252, 192, 191}
\definecolor{mydrawgray}{gray}{0.4}
\definecolor{checkmarkgreen}{rgb}{0,0.6,0} 
\definecolor{crossred}{rgb}{0.9,0,0}   
\definecolor{lightgray}{gray}{0.9}     

\newcommand{\good}[1]{\cellcolor{mygreen!50}#1} 
\newcommand{\poor}[1]{\cellcolor{myred!50}#1}   
\newcommand{\average}[1]{\cellcolor{mygray}#1}  
\newcommand{\myhighlight}[1]{\cellcolor{mygray}#1}
\newcommand{\na}[1]{\text{#1}}

\begin{document}

\title{Argus: A Multi-Agent Sensitive Information Leakage Detection Framework Based on Hierarchical Reference Relationships}
\author{%
Bin Wang$^{1}$,
Hui Li$^{1,*}$,
Liyang Zhang$^{2}$,
Qijia Zhuang$^{2}$}
\author{
Ao Yang$^{1}$,
Dong Zhang$^{3}$,
Xijun Luo$^{3,*}$,
Bing Lin$^{4}$%
}

\affiliation{%
  \institution{$^{1}$Guangdong Provincial Key Laboratory of Ultra High Definition Immersive Media Technology, Shenzhen Graduate School, Peking University \quad
               $^{2}$University of Electronic Science and Technology of China \quad
               $^{3}$Tencent Security Platform Department \quad
               $^{4}$China Unicom (Guangdong) Industrial Internet Co., Ltd}
  \country{} 
}
\email{thebinking66@stu.pku.edu.cn,
       lih64@pkusz.edu.cn,
       {2022090908021,2022090917007}@std.uestc.edu.cn}
\email{jarvisya@stu.pku.edu.cn,
       {zalezhang,junjunluo}@tencnet.com,
       gds-cyhlw@chinaunicom.cn}

\begin{abstract}
Sensitive information leakage in code repositories has emerged as a critical security challenge. Traditional detection methods—relying on regular expressions, fingerprint features, and high-entropy \\ calculations-suffer from high false-positive rates, which not only reduce detection efficiency but also significantly increase the manual screening burden on developers. Recent advances in large language models(LLMs) and multi-agent collaborative architectures have demonstrated remarkable potential in tackling complex tasks, offering a novel technological perspective for sensitive information detection. In response to these challenges, we propose Argus, a multi-agent collaborative framework for detecting sensitive information. Argus employs a three-tier detection mechanism that integrates key content, file context, and project reference relationships to effectively reduce false positives and enhance overall detection accuracy. To comprehensively evaluate Argus in real-world repository environments, we developed two new benchmarks—one to assess genuine leak detection capabilities and another to evaluate false-positive filtering performance. Experimental results show that Argus achieves up to 94.86\% accuracy in leak detection, with a precision of 96.36\%, recall of 94.64\%, and an F1 score of 0.955.Moreover, the analysis of 97 real repositories incurred a total cost of only \$2.21. All code implementations and related datasets are publicly available at \url{https://github.com/TheBinKing/Argus-Guard} for further research and application.
\end{abstract}

\copyrightyear{2026}
\acmYear{2026}
\setcopyright{cc}
\setcctype{by}
\acmConference[ICSE '26]{2026 IEEE/ACM 48th International Conference on Software Engineering}{April 12--18, 2026}{Rio de Janeiro, Brazil}
\acmBooktitle{2026 IEEE/ACM 48th International Conference on Software Engineering (ICSE '26), April 12--18, 2026, Rio de Janeiro, Brazil}
\acmPrice{}
\acmDOI{10.1145/3744916.3773208}
\acmISBN{979-8-4007-2025-3/2026/04}

\ccsdesc[500]{Computing methodologies~Natural language processing}
\ccsdesc[500]{Security and privacy~Software security engineering}

\keywords{sensitive information leakage, code repository security, multi-agent systems, large language models, contextual semantic analysis}

\copyrightyear{2026}
\acmYear{2026}
\setcopyright{cc}
\setcctype{by}
\acmConference[ICSE '26]{2026 IEEE/ACM 48th International Conference on Software Engineering}{April 12--18, 2026}{Rio de Janeiro, Brazil}
\acmBooktitle{2026 IEEE/ACM 48th International Conference on Software Engineering (ICSE '26), April 12--18, 2026, Rio de Janeiro, Brazil}
\acmPrice{}
\acmDOI{10.1145/3744916.3773208}
\acmISBN{979-8-4007-2025-3/2026/04}

\maketitle

\section{INTRODUCTION}

Public code repositories, such as GitHub, have become central platforms for developer collaboration and version control in modern software development. These platforms enable developers to efficiently share code, track issues, and manage versions rigorously, thereby significantly enhancing both development efficiency and code quality. However, their open nature also introduces new security challenges, particularly regarding the management and protection of sensitive information \cite{lazarine2020identifying}. According to monitoring data from GitGuardian \cite{gitguardian2024secrets}, sensitive information leakage incidents on GitHub reached 12.8 million in 2023—a 28\% increase over 2022—with the trend continuing upward. These leaks primarily involve API keys, database credentials, private keys, and other critical data, posing serious risks not only to individual privacy but also to enterprises by exposing them to severe security vulnerabilities and potential economic losses \cite{zhang2023don}.The paper \textit{How Bad Can It Git? Characterizing Secret Leakage in Public GitHub Repositories}  \cite{meli2019bad} discusses the prevalence of secret leakage in open-source Git repositories, highlighting the urgency of addressing this issue.

Current approaches to detecting sensitive information leaks can be broadly classified into two categories. The first comprises rule-based detection tools (e.g., Gitleaks and TruffleHog) that rely on regular expressions and entropy calculations\cite{trunk_gitleaks_vs_trufflehog}. The second involves machine learning methods designed to reduce false positives through model training. However, both approaches have inherent limitations. While rule-based tools offer extensive coverage, some tools have a false positive rate of over 80\% \cite{basak2023comparative}, which substantially undermines their utility. As Chess and McGraw \cite{chess2004static} have noted, “an excessively high false positive rate ultimately leads to 100\% of leaks being overlooked because users will eventually disregard the detection results.” Conversely, machine learning methods \cite{saha2020secrets}, though effective in reducing false positives, lack a deep understanding of code semantics, rendering them less effective in managing complex contextual relationships.

In recent years, the advent of LLMs has opened a new technical pathway for sensitive information detection \cite{10305541}. Compared to traditional methods, LLMs offer superior text comprehension, enabling them to deeply analyze code context and identify potential sensitive information. However, relying solely on LLMs presents challenges: they may struggle to precisely verify key formats and identify placeholders, and their output stability can diminish when processing lengthy texts. To overcome these limitations, the concept of “AI-empowered software engineering” has emerged. This approach leverages multiple AI agents working in collaboration to address complex tasks. The principle of “collaborative AI for SE” involves the coordinated operation of several AI agents, each compensating for the limitations of a single agent when tackling intricate problems. For instance, in tasks such as code review and generation, multi-agent systems have demonstrated significant advantages—such as reducing security vulnerabilities by 13\% \cite{nunez2024autosafecoder} when an LLM responsible for code generation collaborates with agents for static analysis and fuzz testing—while ensuring functional correctness. These findings underscore the potential of a collaborative multi-agent strategy in handling the diverse and highly accurate detection requirements of source code sensitive information.

Motivated by these insights, we propose a multi-agent sensitive information detection framework named Argus. This framework employs a three-tier detection mechanism that integrates key content, file context, and project reference relationships, effectively compensating for the limitations of a single LLM. Each agent focuses on a specific detection task, and through their coordinated efforts, the system achieves stable and precise detection outcomes. Additionally, we have developed a comprehensive evaluation dataset that encompasses common sensitive information scenarios found in open-source projects. Experimental results demonstrate that Argus attains a detection accuracy of 94.86\% on this dataset, significantly outperforming existing methods.

The main contributions of this paper are as follows: 
\begin{enumerate}[leftmargin=*]
    \item We propose a novel three level detection mechanism that offers a comprehensive analysis of sensitive information, providing a fresh perspective on applying LLMs in the field of security detection.
    \item We construct two benchmark datasets based on real-world code repository scenarios, covering a wide range of sensitive information types and usage scenarios, thereby establishing a unified standard for evaluating detection tools. In addition, we verify the validity of the secrets in each repository to mitigate potential security risks. 
    \item We design and implement a multi-agent sensitive information detection framework named Argus, which achieves a precision of 96.36\% and a recall of 94.64\% on the benchmark dataset, significantly outperforming previous baseline tools by efficiently identifying genuine leaks while effectively filtering out false positives.
\end{enumerate}

\section{PROBLEM AND MOTIVATION}

\subsection{Problem Description and Definition of Secret Leak Detection}

In this study, a “secret” refers to sensitive information that appears in plaintext within a code repository without any form of masking or encryption. Such information typically exhibits the following characteristics:

\begin{enumerate}[leftmargin=*]
    \item \textbf{Format Characteristics}: These secrets often have fixed prefixes or distinct character structures. For example, an AWS access key might start with “AKIA”, or an RSA private key may be identified by markers such as “-----BEGIN PRIVATE KEY-----”. Existing literature indicates that rule-based methods primarily target these formatted patterns. 
    \item \textbf{Semantic Relevance}: The content is semantically tied to authentication, authorization, or secure communications and is closely linked to actual business operations. A leak of such information—for instance, an API key from OpenAI—could directly cause service disruptions or financial losses.
\end{enumerate}

Based on these features, this work defines a “secret leak” as the occurrence where any file in a code repository contains plaintext information that meets the definition of a secret and has not been properly masked or encrypted. It is important to note that some texts meeting these characteristics may also be present in a repository; however, if contextual cues or repository indicators make it clear that the secret was intentionally made public by the developer, it should not be considered a secret leak. \textbf{Thus, the core task of this work is to accurately identify and pinpoint unintentional secret disclosures by developers.}

\subsection{Excessive False Positives}
Current methods for detecting secret leaks in code repositories (e.g., TruffleHog, Gitleaks) suffer from severe false positive issues. This not only reduces the practical efficiency of these tools but also significantly increases the manual review burden on developers, effectively rendering a high false positive rate equivalent to low detection accuracy in practice \cite{Ami_2024}. Most existing detection tools rely on custom rules based on regular expressions, fingerprint features, and high-entropy calculations. However, these approaches have clear limitations. For example, many tools mistakenly classify commit hash strings as sensitive information. Similarly, placeholder strings intentionally left by developers (e.g., key templates in the form “sk-xxxxxxxxxxxxxxxxx”) are erroneously flagged as leaks even though they are merely intended to guide the user in entering the actual key.

\subsubsection{General Evaluation}

\begin{table}[h!] 
\centering 
\caption{Leak Data Statistics}
\resizebox{\linewidth}{!}{ 

\begin{tabular}{lccccc} 
\hline 

\textbf{Platform} & \textbf{TE} & \textbf{RL} & \textbf{LR (\%)} & \textbf{RL$>$5} & \textbf{TR} \\

\hline 
\textbf{GitLab} & 1,606,827 & 9,803 & 23.44 & 2,330 & 41,826 \\ 
\textbf{GitHub} & 2,295,293 & 37,149 & 6.21 & 8,287 & 597,933 \\ 
\textbf{Gitee} & 494,247 & 14,750 & 7.27 & 4,512 & 203,012 \\ 
\hline 
\end{tabular} 
\label{tab:leakage_stats} 

}
\begin{minipage}{\linewidth}

\footnotesize\textcolor{gray}{Note:TE = Total Entries, RL = Repositories with Leaks, LR = Leak Repo Ratio, RL$>$5 = Repositories with >5 Leak Entries, TR = Total Repositories.}
\end{minipage}

\end{table}

In our comprehensive evaluation, we employed the actively maintained open-source tool TruffleHog to survey 2,022 mirror backups from GitHub, GitLab and Gitee. Scanning the entire dataset proved prohibitively expensive, so we randomly sampled a large number of repositories to estimate the false positive rate. Given the manpower required to manually verify every detection, we initially treated TruffleHog’s outputs as ground truth. The detailed results appear in Table~\ref{tab:leakage_stats}.

Our analysis shows that over 7.3\% of repositories contained at least one reported leak, rising to 23.44\% on GitLab, and that approximately 5.57\% of repositories reported more than five leaks. Structured data files (\texttt{.csv}, \texttt{.json}) accounted for roughly 520 000 detections while document files (\texttt{.md}, \texttt{.txt}) comprised about 10\% of all findings. Notably, repositories reporting more than 50 leaks contributed 75.69\% of the total detection volume.

A closer examination of TruffleHog’s outputs revealed a high prevalence of false positives concentrated in just a few rules. For example, entries flagged by the “Github” and “Gitlab” rules made up 72.4\% of all detections, yet many of these corresponded to commit hashes or default configuration files misclassified as secrets. Likewise, the “JDBC” and “URI” rules proved overly broad, frequently tagging test data and boilerplate as sensitive. On Gitee, such spurious entries totaled around 26000, representing 5\% of all flags. We found that these noisy files typically feature templated structures, highly repetitive fields and rigid formatting, causing the same rule to trigger repeatedly across multiple projects. This systematic amplification of false alarms places a heavy burden on downstream analysis.

To validate our false positive assessment, we randomly sampled 2000 entries from TruffleHog’s full output. Two independent annotators with security expertise reviewed each record, classifying it as a genuine secret or a false positive based on contextual semantics, structural patterns and known non-sensitive markers. A third reviewer resolved any disagreements. The final annotations showed that fewer than 3.4\% of sampled records represented genuine leaks; the vast majority consisted of default values, placeholders, debug information or highly repetitive strings. These findings confirm that while secret leaks are indeed widespread, false positives are pervasive .

\subsubsection{False Positive Experiment}

\begin{table}[h]
\centering
\caption{False Positive Statistics with Version Information}
\renewcommand{\arraystretch}{1.3} 
\resizebox{0.8\linewidth}{!}{ 
\begin{tabular}{llccccc}
\toprule
\textbf{Repository} & \textbf{Version} & \textbf{TH} & \textbf{GL} & \textbf{ST} & \textbf{WP} \\ 
 
\midrule
\textbf{moby}        & c710b88  & 83  & 181  & (148, 11, 0)   & 73  \\ 
\textbf{kubernetes}  & 9253c9b  & 142  & 306  & (110, 66, 19)  & 27  \\ 
\textbf{bitcoin}     & bf03c45  & 3    & 71   & (2, 2, 102)    & 12  \\ 
\textbf{neovim}      & 8b98642  & 5    & 7    & (2, 0, 0)      & 3   \\ 
\textbf{webpack}     & 3612d36  & 17   & 1    & (1, 1, 0)      & 4   \\ 
\textbf{spring-boot} & 8964203  & 56   & 26   & (28, 11, 11)   & 2   \\ 
\textbf{fastapi}     & 113da5b  & 1    & 28   & (45, 0, 0)     & 1   \\ 
\textbf{pandas}      & 0691c5c  & 2    & 1    & (1, 2, 0)      & 4   \\ 
\textbf{vue}         & 13f4e7d  & 56   & 1    & (1, 0, 0)      & 3   \\ 
\textbf{transformers}& 5d7739f  & 5    & 13   & (0, 66, 8)     & 2   \\ 
\bottomrule
\end{tabular}
}
\label{table:false_positive}
\begin{minipage}{\linewidth}

\footnotesize\textcolor{gray}{Note: TH = TruffleHog,GL = Gitleaks,ST = SpectralOps,WP = Whispers,The values of ST represent the number of detections with (high, mid, low) severity levels.}
\end{minipage}
\end{table}


\begin{figure*}[htbp]
    \centering
    \begin{minipage}{0.66\textwidth} 
        \subfloat[Composition of dataset categories]{
            \includegraphics[width=0.48\linewidth]{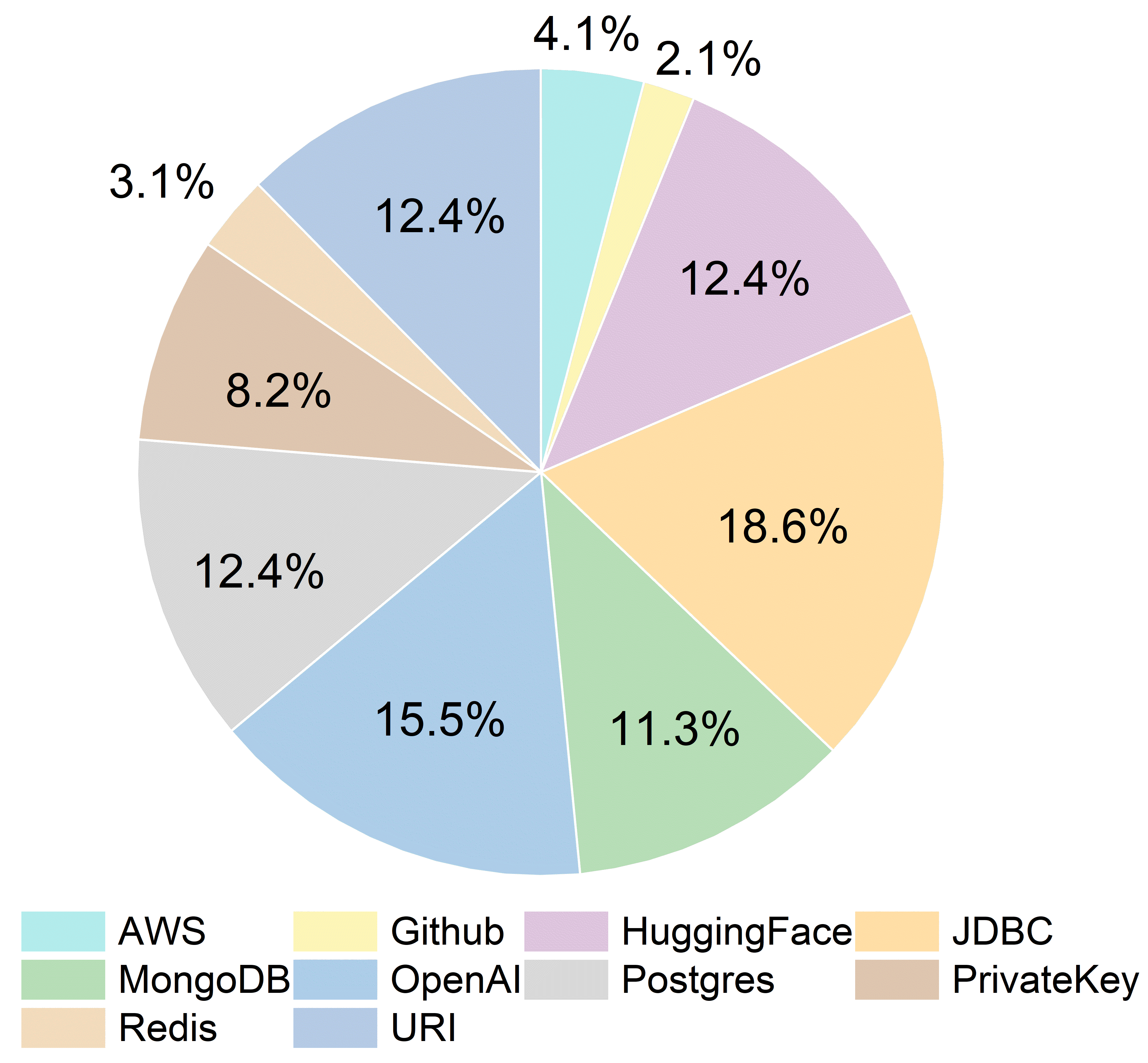}
        }
        \hspace{0.01\linewidth}
        \subfloat[Composition of Dataset Language]{
            \begin{tikzpicture}  
                \begin{scope}[yshift=-10pt]
                \node (img) at (0,-1){\includegraphics[width=0.4\linewidth]{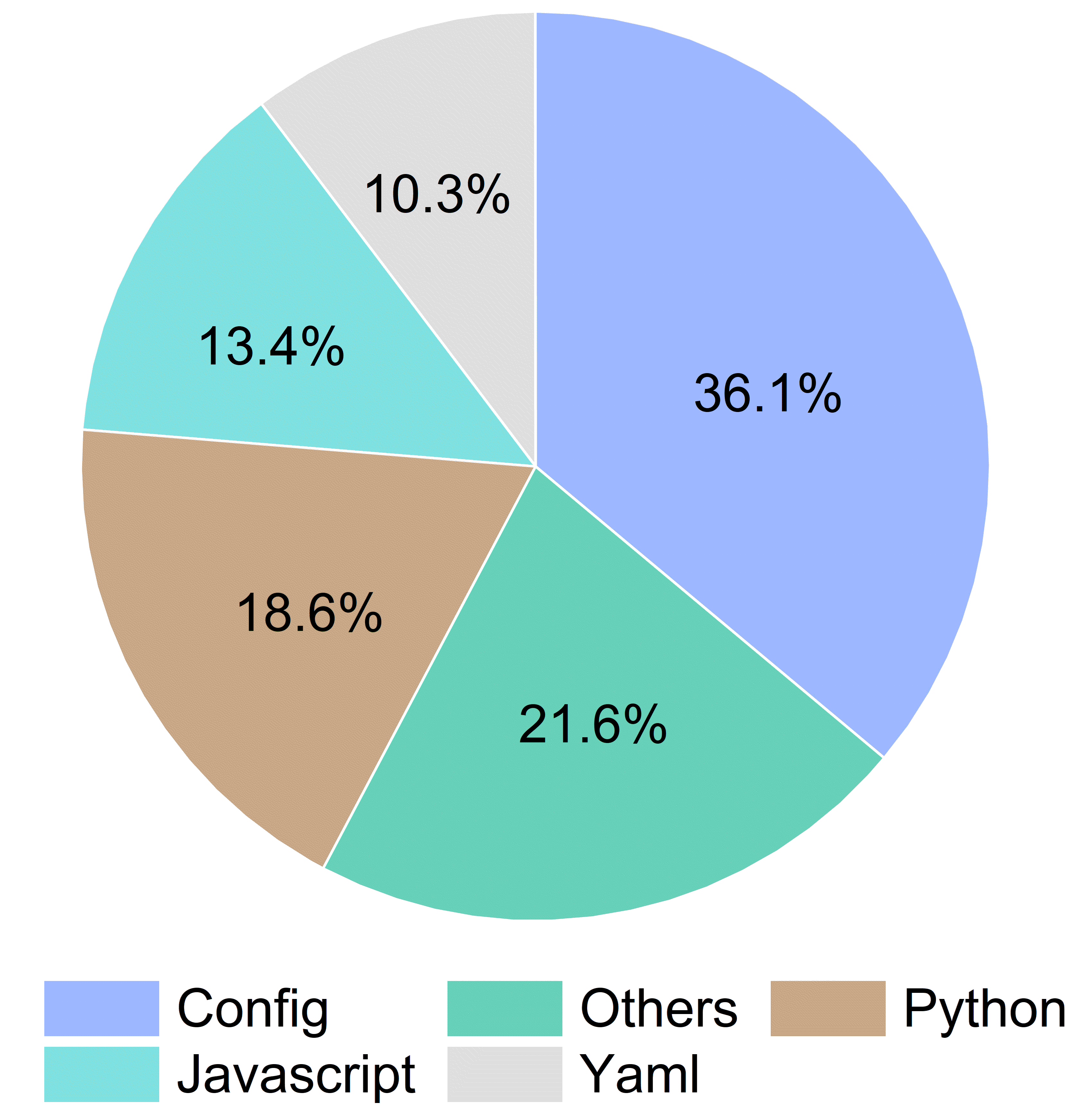}};
                \end{scope}
                \draw[dashed, thick] (0.4,1) --(3.4,0)node[right]{};
                \draw[dashed, thick] (-0.32,-2.9) -- (3.3,-3.9)node[right]{};
            \end{tikzpicture}
        }
    \end{minipage}
    \hfill 
    \begin{minipage}{0.33\textwidth} 
        \centering
        \scriptsize
        \renewcommand{\arraystretch}{1.1} 
        \setlength{\tabcolsep}{4pt} 
        \captionof{table}{Composition of Config and Others}
        \label{tab:config_others-combined_horizontal_tight}
        \begin{tabular}{@{}p{0.22\linewidth}p{0.09\linewidth}p{0.09\linewidth} @{\hspace{1em}} p{0.22\linewidth}p{0.09\linewidth}p{0.09\linewidth}@{}}
        \toprule
        \multicolumn{3}{c}{\textbf{Config}} & \multicolumn{3}{c}{\textbf{Others}} \\
        \cmidrule(r){1-3} \cmidrule(l){4-6} 
        Subcategory & Total & Prop. & Subcategory & Total & Prop. \\
        \midrule
        Env         & 9  & 25.71\% & Java        & 5  & 23.81\% \\
        Json        & 5  & 14.29\% & CS          & 2  & 9.52\% \\
        Properties  & 4  & 11.43\% & Dockerfile  & 2  & 9.52\% \\
        Ipynb       & 3  & 8.57\%  & Shell       & 2  & 9.52\% \\
        Markdown    & 3  & 8.57\%  & Typescript  & 2  & 9.52\% \\
        Key         & 3  & 8.57\%  & PHP         & 2  & 9.52\% \\
        Git         & 2  & 5.71\%  & C           & 2  & 9.52\% \\
        Data        & 2  & 5.71\%  & Gradle      & 1  & 4.76\% \\
        Pem         & 2  & 5.71\%  & Html        & 1  & 4.76\% \\
        Txt         & 1  & 2.86\%  & TCL         & 1  & 4.76\% \\
        Conf        & 1  & 2.86\%  & CPP         & 1  & 4.76\% \\
        \bottomrule
        \end{tabular}
    \end{minipage}
    \caption{Composition of Dataset and Subcategories}
    \label{fig:dataset-comparison}
\end{figure*}

To assess the limitations of current sensitive information detection methods, we selected 10 high-star repositories from GitHub and analyzed the false positive counts using four tools: TruffleHog, Gitleaks, SpectralOps, and Whispers. The experimental results indicate that TruffleHog generated 370 false positives. Although its deep scan strategy helps in effectively filtering out high-entropy strings, there is still room for improvement in handling complex encodings and boundary cases. Gitleaks, despite offering broad detection coverage, produced 635 false positives due to overly broad rule settings that led to numerous false positives in test code and template files, thereby increasing the manual review burden. SpectralOps reported the highest number of false positives. Although it categorizes the results into high, medium, and low risk to provide developers with a prioritization reference, its reliance on machine learning and context analysis has not sufficiently reduced false positives from non-sensitive content. In contrast, Whispers generated only 131 false positives, a lower count primarily attributable to its limited detection scope (focusing solely on hard-coded files) and restricted language support (limited to JavaScript, Java, Go, and PHP)(see Table 2 for details).

Overall, although the actual occurrence of sensitive information leaks in these high-star repositories is relatively low, the prevalent issue of excessive false positives not only increases the manual review workload for developers but also risks overlooking genuine sensitive information. To enhance detection efficiency and security, future strategies must aim to reduce false positives further—through the incorporation of context analysis and dynamic rule adjustment—while maintaining a broad coverage.

\section{DATASETS}
\label{sec:datasets}

Current datasets in the secret leak detection domain exhibit several shortcomings. First, most datasets focus exclusively on a single type of secret (e.g., keys or credentials), resulting in a significant gap between the collected data and what is observed in real-world code repositories \cite{lounici2021optimizing}. Second, these datasets generally lack hierarchical grading and detailed categorization of sensitive information, making it difficult to thoroughly evaluate the distinct characteristics and risks associated with various types of secrets. Moreover, due to the variable quality of projects on GitHub, many existing datasets inadvertently include a large number of low-quality or inactive projects, which introduces sample bias. Lastly, even when some datasets are sourced from reputable repositories, the secrets contained therein may still be active, thereby posing additional sensitivity and security risks \cite{feng2022automated}  \cite{basak2023secretbench}.

To address the limitations found in existing datasets—such as limited secret types, lack of hierarchical annotation, and insufficient validation—we construct two new datasets: CommonLeak and TrustedFalseSecrets. CommonLeak is based on a TruffleHog scan of a GitHub snapshot from June 2022. We manually selected 97 representative projects covering ten common secret types, including AWS, GitHub, Huggingface, JDBC, MongoDB, OpenAI, PostgreSQL, PrivateKey, Redis, and URI. Each candidate was reviewed by two independent annotators using context and semantics to distinguish real secrets from false positives. Disagreements were resolved by a third reviewer. All confirmed true leaks were deactivated for safe release. The final dataset contains 57 true positives and 40 false positives. Details are shown in Figure~\ref{fig:dataset-comparison} and Table~\ref{tab:config_others-combined_horizontal_tight}. TrustedFalseSecrets focuses on representative false positives; we curated 20 typical cases from ten well-maintained open-source repositories to illustrate common misclassifications made by regex-based tools. This dataset offers a clean benchmark for evaluating false positive mitigation techniques (see Table~\ref{tab:famouse_check}).

\section{METHODOLOGY}
In this section, we present the design of Argus. From a methodological perspective, Argus employs a three-level analysis framework to assess secrets, leveraging a multi-agent collaboration mechanism to distribute and coordinate tasks. Additionally, it utilizes a shared memory pool to record intermediate processes and facilitate information sharing among agents.

\begin{figure*}[tbp]  
\centering  
\includegraphics[width=0.85\textwidth]{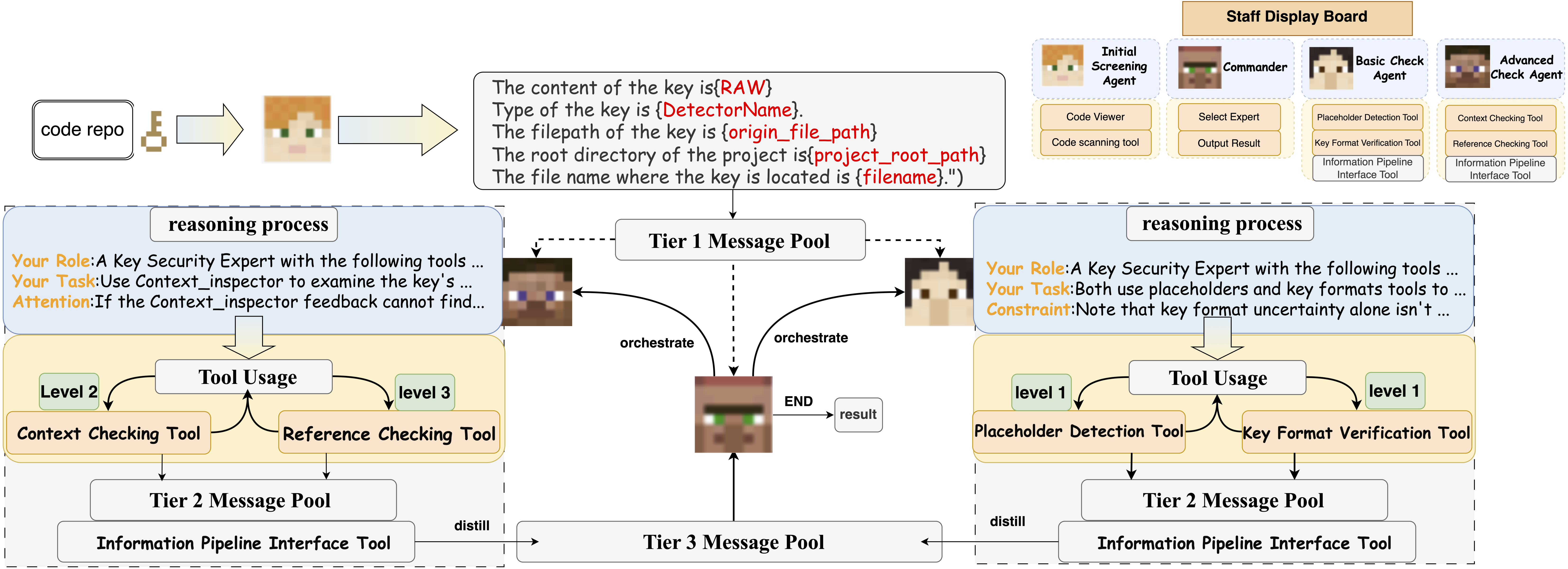}  
\caption{Overview of the Argus framework and its operational flow}  
\label{fig:framework} 
\end{figure*}

\subsection{Three-Level Contextual Semantic Analysis}
Traditional tools for scanning code repositories for secrets are frequently overwhelmed by false positives. To address this challenge, we propose a detection method based on three-tier contextual semantic analysis, implemented through a multi-agent system that automates decision-making. This approach decomposes the secret detection task into three interconnected layers: the analysis of intrinsic key features, the semantic interpretation of its immediate context, and the examination of project-level reference relationships. Together, these layers form a hierarchical, traceable, and interpretable detection process (As shown in Figure \ref{fig:framework}).

\subsubsection{Level 1: Analysis of Intrinsic Semantics}

At this initial level, the focus is solely on the secret’s own features. The goal is to rapidly dismiss obvious false positives by inspecting characteristics such as readability, placeholder usage, and adherence to specific key formats. False positives at this stage generally fall into three categories:

\begin{enumerate}[leftmargin=*]

\item \textbf{Readable Keys}: For example, a string like \texttt{https://readonly:\allowbreak readonly@www.pauldreik.se} is semantically clear and lacks the high entropy or specific structure expected of a genuine key. Traditional tools that rely solely on entropy and regex matching often fail to properly filter such “readable” pseudo-keys, whereas LLMs can use their semantic analysis capabilities to recognize these non-genuine characteristics.

\item \textbf{Keys with Placeholders}: For instance, \texttt{mongodb://username: \allowbreak password@server} appears in documentation (e.g., Markdown files) as an example, using fixed placeholders like \texttt{username} or \texttt{password}. By analyzing large-scale data, we have identified a set of common placeholders. Our placeholder detection tool checks for these markers within the key, thereby flagging such cases as likely false positives.

\item \textbf{Keys Not Conforming to Specific Formats}: For example, \texttt{jdbc:postgresql:// \allowbreak my.domain.com} might be a truncated version of a legitimate key format, omitting the required username and password. Traditional regex-based detection does not differentiate between valid and invalid formats across key types. To remedy this, we have designed precise regex patterns for major key types (e.g., AWS, JDBC, MongoDB) to verify compliance with their expected formats.
\end{enumerate}

Implementation-wise, each tool comprises an LLM and a specific function. For example, a key format checker integrates a GPT-4o model with a custom prompt and a regular expression matching function. The function’s parameters, return values, and usage are embedded in the prompt to enable the LLM to accurately utilize the tool and provide correct feedback based on its detection results.Each agent consists of an LLM and multiple tools. For instance, a preliminary inspection agent includes an LLM with a custom prompt and tools such as placeholder and key format checkers. The definitions of these tools are incorporated into the prompt to guide the agent in selecting appropriate tools for secret inspection.
Unlike traditional secret detection tools, the detection results (e.g., regular expression matches) are not treated as final conclusions but rather as evidence for the LLM’s judgment. The LLM performs secondary analysis and inference, considering the actual characteristics of the key, to infer whether the key is genuine or merely resembles one.

Implementation-wise, the relevant tools are encapsulated within individual agents. Instead of directly using regex match results as the final verdict, these results are provided to an LLM, which combines the clues with the key’s intrinsic features to infer whether the key is genuine or merely resembles one.

\subsubsection{Level 2: Semantic Analysis of the Secret’s Immediate Context}

At the second level, the focus shifts to cases where a key, while appearing authentic on its own, is intended solely for demonstration, teaching, or testing. Because such keys exhibit all the intrinsic characteristics of genuine secrets, a standalone analysis is insufficient. Instead, the surrounding context must be examined.

For instance, a document might include a  \texttt{SECRET\_ACCESS\_KEY}  example accompanied by explanatory text clarifying its instructional purpose. Traditional tools might simply flag the key as a risk, but our multi-agent system features an advanced context analysis module. This module scrutinizes annotations, comments, and nearby narrative cues to determine if the key is merely illustrative rather than operational.

\subsubsection{Level 3: Global Reference Analysis at the Project Level}

In cases where keys are embedded as standalone files (e.g., RSA private keys or certificates) and display all the attributes of genuine secrets, relying solely on intrinsic feature analysis or immediate contextual evaluation may not yield definitive results. To address this, Level 3 detection examines the key’s role and its relationships within the entire project. Below is an illustration of the Level 3 detection process using an RSA private key inspection as an example:

\begin{enumerate}[leftmargin=*]

\item \textbf{Initial Discovery}: The scanning tool detects a file matching the RSA private key format. Since it does not trigger obvious false positive conditions in tiers one or two, its authenticity remains undetermined.

\item \textbf{Reference Path Check}: The advanced module retrieves the file’s reference location within the project. For example:

\begin{mdframed}[backgroundcolor=lightgray] 
\textbf{The RSA private key is referenced in the file:} \\
\texttt{\detokenize{final_dataset\PrivateKey\...\pay.py}}
\end{mdframed}

This suggests the key is likely utilized by a functional module.

\item \textbf{Contextual Analysis of the Reference}: A further examination of the code in \texttt{pay.py} shows that the key file is read and assigned to a variable (e.g., \texttt{app\_private\_key\_string}) in conjunction with terms like \texttt{alipay\_public\_key\_string}, indicating its role in genuine payment or encryption operations.

\item \textbf{Final Determination}: Lacking any indicators that the key is used for testing or demonstration, and given its active usage in core functionalities, the system concludes that it is a genuine secret leak.
\end{enumerate}

Overall, Level three focuses on project-level usage and reference relationships. If a key cannot be ruled out as a false positive via intrinsic or contextual analyses, examining its practical deployment (through reference paths, function calls, or file dependencies) often yields the final determination: if it is employed in production, it is treated as a genuine leak requiring immediate remediation.

\subsection{Role Specialization}

In our multi-agent system, we first designate an initial screening agent to locate high-entropy or feature-based secret candidates. Next, a \textbf{Commander} acts as the ultimate decision-maker, delegating tasks to two specialized roles: the \textbf{Basic Check Agent} and the \textbf{Advanced Check Agent}. Each role is equipped with specific tool and functional capabilities, working together to determine the authenticity of a secret.

\subsubsection{Initial Screening Agent}

Argus employs a TruffleHog-based wrapper as its initial screening agent. This agent scans the designated code repository to produce a preliminary set of candidate secrets. Although this stage—similar to existing detection logic—often produces numerous false positives, it effectively narrows down the scope and reduces the token usage required by subsequent LLM multi-agent processes.

\subsubsection{Commander}

The Commander serves as the system’s central scheduling node. Its primary responsibility is to select the appropriate check agent based on the initial information and system prompts, and to integrate the results from each stage to make a final decision. The process involves both preliminary screening by the Basic Check Agent and, when necessary, deeper analysis by the Advanced Check Agent.

For suspected key detection, the Commander receives an input sample \(x \in \mathcal{X}\) containing all key-related information (e.g., key content, file and project paths, filenames). At the same time, the system defines a set of prompt templates  \(\mathcal{P}_{\text{prompt}}\) used both to generate decisions \(p_{\text{gen}}\) and to instruct check
agent\ \(p_{\text{agent}}\), Under this framework, the Commander is formalized as a mapping function:

\begin{equation}
\mathcal{M}_{\text{super}}: \left(\mathcal{X} \parallel \mathcal{P}_{\text{prompt}}\right) \rightarrow Y.
\label{eq:Commander_mapping_function}
\end{equation}

This function takes the input sample and prompt templates to output an element  \(y \in Y\) from the decision space. The output may either be a direct determination of the suspected key or an indication to invoke further analysis.

To implement a tiered detection process, Argus defines two check agent, forming the agent set  \(\mathcal{T}=\{\text{Basic},\text{Advanced}\}\) Each agent acts as an operator that, based on the current shared information pool \(\mathcal{M}\) and the prompt \(p_{\text{agent}}\), produces detection results:

\begin{equation}
\begin{aligned}
\mathcal{T}_t &: \left(\mathcal{M} \parallel \mathcal{P}_{\text{prompt}}\right) \rightarrow \mathcal{R}.
\end{aligned}
\label{eq:detection_results}
\end{equation}

where \(\mathcal{R}\) is the set of detection outcomes. The Commander uses a path selection function:

\begin{equation}
\begin{aligned}
\mathcal{P} &: Y \rightarrow \mathcal{T} \cup \{\text{END}\}.
\end{aligned}
\label{eq:path_selection_function}
\end{equation}

to decide whether to invoke a check agent or terminate the process. If \(\mathcal{P}(y) = \text{END}\), the system outputs the final decision.

Throughout this process, intermediate results are stored in a shared information pool \(\mathcal{M}\). Each detection result \(R\) is merged into \(\mathcal{M}\) via an accumulation operation \(\oplus\), ensuring coherent and robust decision-making based on historical data.

The following algorithm summarizes the Commander’s scheduling and decision-making process:

\begin{algorithm}[H]
\caption{Commander Scheduling and Decision Process}
\begin{algorithmic}[1]
\Require Input sample \(x \in \mathcal{X}\) \\
\hspace{12mm} Decision mapping \(\mathcal{M}_{\text{super}}\): \((\mathcal{X} \parallel \mathcal{P}_{\text{prompt}}) \to Y\) \\
\hspace{12mm} Path selection function \(\mathcal{P}: Y \to \mathcal{T} \cup \{\text{END}\}\) \\
\hspace{12mm} Prompt templates \(p_{\text{gen}}, p_{\text{agent}} \in \mathcal{P}_{\text{prompt}}\)

\State Initialize information pool: \(\mathcal{M}_0 \gets \varnothing\)
\State Compute initial decision: \(y_0 \gets \mathcal{M}_{\text{super}}\bigl(x \parallel p_{\text{gen}}\bigr)\)
\State Select check agent: \(t_0 \gets \mathcal{P}(y_0)\)
\If{\(t_0 = \text{END}\)}
    \State \Return \(\text{Final Decision from } y_0\) \Comment{Terminate immediately}
\EndIf

\State Obtain preliminary detection result: \(R_0 \gets \mathcal{T}_{t_0}\bigl(\mathcal{M}_0 \parallel p_{\text{agent}}\bigr)\)
\State Update information pool: \(\mathcal{M}_1 \gets \mathcal{M}_0 \oplus R_0\)

\For{\(k = 0,1,2,\dots\)}
    \State Compute new decision output: \(y_{k+1} \gets \mathcal{M}_{\text{super}}\bigl(R_k \parallel p_{\text{gen}}\bigr)\)
    \State Select next check agent: \(t_{k+1} \gets \mathcal{P}(y_{k+1})\)
    \If{\(t_{k+1} = \text{END}\)}
        \State \Return \(\text{Final Decision from } y_{k+1}\)
    \Else
        \State Obtain new detection result: \(R_{k+1} \gets \mathcal{T}_{t_{k+1}}\bigl(\mathcal{M}_{k+1} \parallel p_{\text{agent}}\bigr)\)
        \State Update information pool: \(\mathcal{M}_{k+2} \gets \mathcal{M}_{k+1} \oplus R_{k+1}\)
    \EndIf
\EndFor
\end{algorithmic}
\end{algorithm}

\subsubsection{Basic Check Agent}

Within Argus, the Basic Check Agent acts as the first line of defense, quickly verifying candidate keys with low overhead and providing reliable preprocessed information for further analysis.

Specifically, the Basic Check Agent relies on two core tools:

\begin{itemize}[leftmargin=*]
    \item \textbf{Key Format Verification Tool.} This tool evaluates whether a key adheres to the inherent format characteristics expected of its type. For example, AWS keys typically start with ``AKIA'' followed by a fixed-length sequence, while RSA private keys follow a specific multiline structure. This allows the system to swiftly flag keys that do not conform to standard formats.
    
    \item \textbf{Placeholder Detection Tool.} This tool identifies common placeholders (e.g., “username,” “password,” “host”) that suggest the data is merely an example or intended for testing. By matching these common indicators, the system can further filter out false positives.
\end{itemize}

To ensure seamless transmission of detection information to the upper decision-making module, the Basic Check Agent incorporates an \textbf{Information Pipeline Interface Tool}. This tool standardizes and organizes the results from both verification processes and outputs them to the shared memory pool, thereby providing the Commander with consistent, high-quality input.

\subsubsection{Advanced Check Agent}

For keys that remain ambiguous after the basic screening, the Advanced Check Agent conducts in-depth analysis and verification. Drawing inspiration from expert systems and deep semantic analysis, this agent performs validation at both file and project levels, leveraging richer contextual data and cross-document reference analysis to determine the key’s true purpose.

The Advanced Check Agent uses two main tools:

\begin{itemize}[leftmargin=*]
    \item \textbf{Context Checking Tool.} This tool employs the semantic understanding of LLMs to analyze the context surrounding a key—especially in documents like Markdown—thus discerning whether the key is merely an illustrative example or intended for production use.
    \item \textbf{Reference Checking Tool.} This tool traces the reference paths of a key file by identifying and analyzing the higher-level files that reference it. This process not only uncovers the key’s actual usage but also distinguishes between educational examples and genuine secrets.
\end{itemize}

Similar to the Basic Check Agent, the Advanced Check Agent is equipped with an \textbf{Information Pipeline Interface Tool} that standardizes its outputs and transmits them via the shared memory pool for the Commander’s final decision-making.

\subsection{Shared Memory Pool}
Efficient information sharing is crucial for reducing communication overhead and enhancing coordination within a multi-agent system. Traditional one-to-one messaging can lead to complex topologies and inefficiencies. Inspired by the global shared message pool design in the MetaGPT framework, we have developed a three-tier shared memory pool structure to meet the diverse needs of the agents:

\begin{itemize}[leftmargin=*]
    \item \textbf{first-tier message pool}: Stores all secret-related data associated with keys, accessible by every agent to extract necessary details for their tasks.
    \item \textbf{second-tier message pool}: Records detailed processes and feedback from the Basic and Advanced Check Agents during tool invocations. This raw, comprehensive data serves as a rich basis for further refinement.
    \item \textbf{third-tier message pool}: Contains highly structured, refined results produced by summarizing and optimizing the second-tier data. These optimized outcomes—including both basic and advanced check conclusions—are ultimately used by the Commander to achieve efficient and accurate final decisions.
\end{itemize}
This three-tier design not only ensures synchronization of information across agents but also significantly enhances communication efficiency and data reuse, providing robust support for complex key detection tasks.

\subsection{Case Study: Demonstrating Argus in Action}
\begin{figure*}[tbp] 
\centering  
\includegraphics[width=0.85\textwidth]{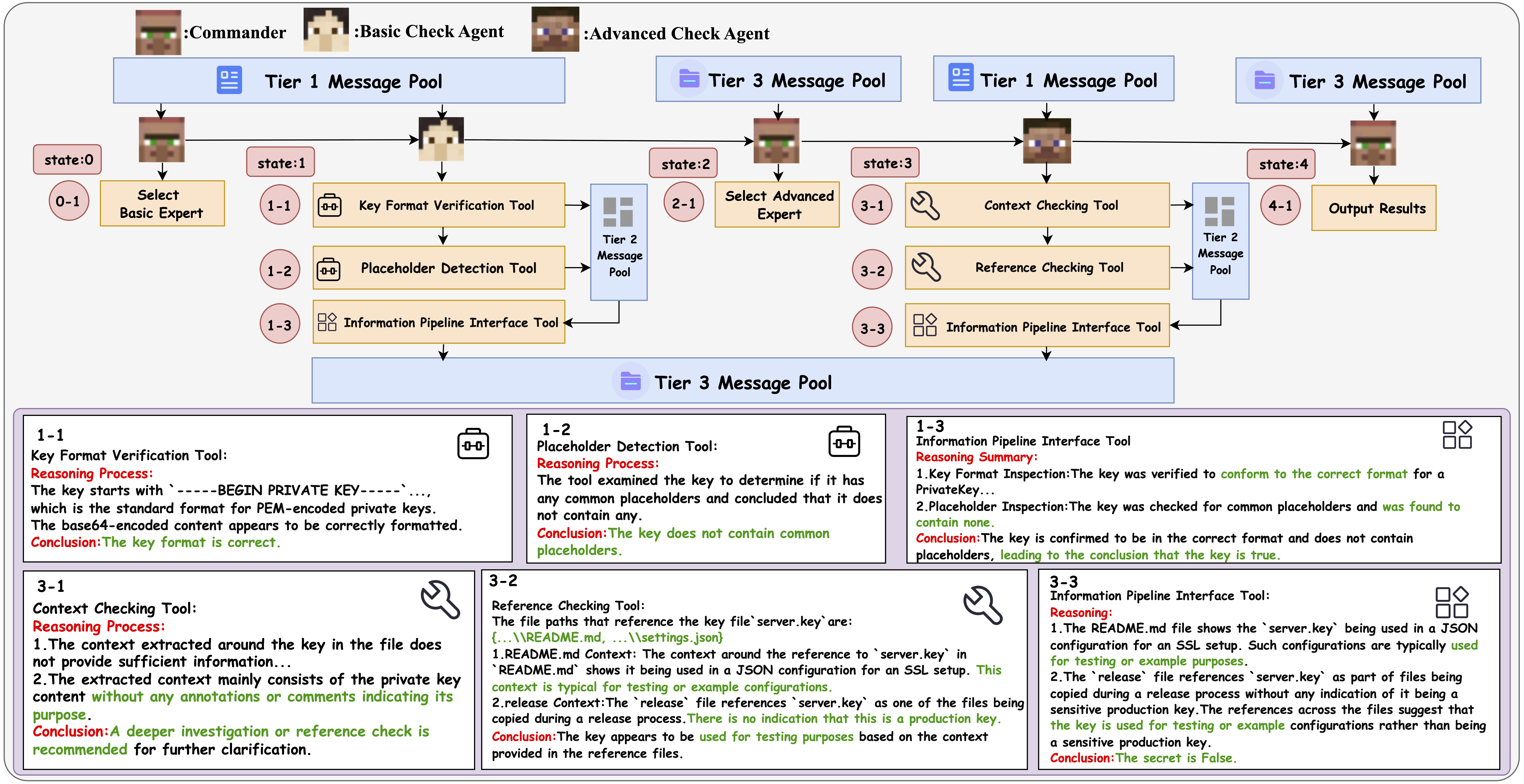} 
\caption{Illustration of the Argus workflow: multi-level analysis and agent collaboration}  
\label{fig:flowcase}  
\end{figure*}

In the following example ,as shown in Figure \ref{fig:flowcase},we demonstrate Argus’s overall workflow using a real-world scenario. First, after an initial detection and organization by the Initial Screening Agent, the secret information to be analyzed is stored in the first-tier message pool. The Commander then reviews this data and assigns a Basic Agent to carry out the initial inspection. The Basic Agent retrieves the secret information from the first-tier pool, uses a key format detection tool and a placeholder detection tool, and stores the resulting outputs in the second-tier message pool. Next, the Information Pipeline Interface Tool integrates these second-tier results to form a preliminary conclusion, which is then stored in the third-tier message pool.

The Commander then evaluates the initial conclusions in the third-tier message pool to decide whether to deploy an Advanced Agent for further analysis. If the Advanced Agent’s context checking tool identifies insufficient contextual information, a global reference check is performed. The outcome of this check is also stored in the third-tier message pool. Finally, based on the advanced inspection results in the third-tier message pool, the Commander determines whether the secret in question is genuine.

\section{SECRET LEAKAGE ANALYSIS}

\label{sec:Analysis}

The experiments focus on three core research questions (RQs):

\begin{itemize}[leftmargin=*]
    \item \textbf{RQ1:} How effective is Argus in detecting secrets? This evaluates Argus's ability to detect sensitive information leaks in real-world projects.
    \item \textbf{RQ2:} How does Argus perform in the analysis of false positives in secret detection? This verifies Argus's performance regarding false positives.
    \item \textbf{RQ3:} What is the system overhead incurred by Argus during detection?
\end{itemize}

\textbf{Evaluation Datasets.} To avoid potential security risks and ethical concerns, we employ two rigorously reviewed datasets—\texttt{CommonLeak} and \texttt{TrustedFalse\allowbreak Secrets}—to evaluate the model's secret detection and false positive identification capabilities. The detailed dataset construction can be found in Section~\ref{sec:datasets}.

\textbf{Baseline.} This study selects a diverse set of LLMs and traditional secret detection tools as baseline comparators, categorized into two main types: regex-driven and AI-driven methods. These tools are highly representative in the field of secret detection. Detailed information can be found in Table~\ref{table:baseline_tools}.

\begin{table*}[h]
\centering
\caption{Descriptions of Baseline Tools (with abbreviations used in this paper, e.g., TruffleHog is denoted as TRU).}
\resizebox{\textwidth}{!}{
\begin{tabular}{@{}llccp{6cm}@{}}
\toprule
\textbf{Type} & \textbf{Tool Name} & \textbf{Version} & \textbf{Tool Description} \\ 
\midrule
\multirow{4}{*}{\textbf{Regex-driven}} 
    & \textbf{TruffleHog(TRU)}\cite{trufflehog_github}  & v3.86.1 & Scans repositories for high-entropy strings, secrets, and key leaks using a set of regular expressions. \\
    & \textbf{Gitleaks(GL)}\cite{gitleaks_github}    & v8.21.4 & Detects API keys, tokens, and other secrets in Git repositories and files. \\
    & \textbf{Whispers(WS)}\cite{whispers_github}    & v1.5.3 & A static code analysis tool designed to parse various common data formats and identify hardcoded credentials. \\
    & \textbf{git-secrets(GS)}\cite{gitsecrets_github} & v1.3.0 & Prevents secret leaks by enforcing user-defined regular expressions via Git hooks. \\
\midrule
\multirow{4}{*}{\textbf{AI-driven}} 
    & \textbf{SpectralOps(ST)}\cite{spectralops_website} & v1.10.233 & Uses AI technology to automatically monitor, classify, and protect assets throughout the CI/CD pipeline. \\
    & \textbf{gpt-4o(GD)} & 2024-08-06 & Multimodal AI model capable of processing and generating text, audio, and images\\
    & \textbf{gpt-4o-CoT(GC)} & 2024-08-06 & Chain-of-Thought (CoT) prompting guides LLM to generate intermediate reasoning steps, enhancing their ability to solve complex problems.\\
    & \textbf{gpt-3.5-turbo(GT)} & 2023-01-25 &optimized version of OpenAI's GPT-3.5 model, designed to provide faster inference speeds, higher throughput, and improved resource efficiency\\
\bottomrule
\end{tabular}
}
\label{table:baseline_tools}
\end{table*}

\subsection{Secret Leakage Analysis}

This section mainly presents a comparative study between Argus and state-of-the-art baselines on the \texttt{CommonLeak} dataset, in order to evaluate Argus's ability to detect sensitive information leaks in real-world projects.  The summary of the experimental comparison is illustrated in the Overall Accuracy subplot of Figure \ref{fig:acc-compare-program-language}.

Argus achieves an overall accuracy of 94.86\% on the entire dataset, which is higher than that of the GPT-4o model (75.3\%), the GPT-4o model with the CoT approach (67.0\%), and the GPT-3.5-Turbo model (59.8\%). It is also significantly higher than traditional tools such as TruffleHog (57.7\%), Git-Secrets (63.9\%), Gitleaks (61.9\%), Whispter (53.6\%), and Spectralops (69.1\%), demonstrating the superiority of the multi-agent system in verifying the authenticity of secrets over both large models and traditional detection tools.

Since projects with different types of secret leaks and programming languages often involve distinct scenarios, we evaluate the secret leakage detection performance from two perspectives: the primary language of the project and the type of secret leakage.

\subsubsection{Comparison results of language category detection}

For different programming language scenarios, the data distribution is illustrated in the right diagram of Figure~\ref{fig:dataset-comparison}. The dataset primarily consists of projects written in three major programming languages: Python, YAML, and JavaScript, which collectively represent the dominant portion of the dataset.In the language classification results, 35 files containing leaked secrets belong to configuration files whose contexts are not directly related to other programming languages. Therefore, we label these files as \textbf{Config} and further categorize them into 11 classes based on their file extensions. Due to the smaller number of files in individual language categories, all other languages are collectively labeled as \textbf{Other Languages}, which include 11 programming languages: C, CPP, CS, TCL, PHP, Typescript, HTML, Java, Gradle, Shell, and Dockfile. For details on the dataset composition, please refer to Figure~\ref{fig:dataset-comparison} and Table~\ref{tab:config_others-combined_horizontal_tight}.

\begin{figure*}
  \centering
  \begin{tikzpicture}
    \begin{groupplot}[
      height=3.1cm, width=6.5cm,
      /pgf/bar width=0.25cm,
      xmin=-1.5, xmax=9.5,
      axis x line*=bottom, axis y line*=left, enlarge x limits=false,
      xtick={-0.9,0.9,2,3,4,5.6,6.8,7.85,9},
      xticklabels={Argus, GD, GC, GT, ST, TRU, GS, GL, WS},
      xticklabel style={yshift=-0.3mm, font=\scriptsize, align=center},
      ybar=3.8pt, clip=false,
      ymin=0, ymax=100,
      ytick={0, 25, 50, 75, 100},
      yticklabels={0, 25, 50, 75, 100},
      ymajorgrids, major grid style={draw=black!20},
      tick align=inside,
      yticklabel style={font=\footnotesize},
      tickwidth=0pt,
      y axis line style={opacity=0},
      group style={group size=3 by 2, horizontal sep=25pt, vertical sep=50pt},
    ]

    \nextgroupplot[ylabel={Score}, title={Config}]
    \addplot [fill=mygreen] coordinates {(0,100)}; 
    \addplot [fill=mygray] coordinates {(1,68.57) (2,65.71) (3,60) (4,55.88)};

    \addplot [fill=myred] coordinates {(5,58.82) (6,58.82) (7,38.24) (8,55.88)};

    \nextgroupplot[title={Yaml}]
    \addplot [fill=mygreen] coordinates {(0,90)};
    \addplot [fill=mygray] coordinates {(1,50) (2,50) (3,30) (4,60)}; 
    \addplot [fill=myred] coordinates {(5,60) (6,60) (7,80) (8,60)};

    \nextgroupplot[title={Python}]
    \addplot [fill=mygreen] coordinates {(0,100)};
    \addplot [fill=mygray] coordinates {(1,83.33) (2,83.33) (3,83.33) (4,89.47)}; 
    \addplot [fill=myred] coordinates {(5,84.21) (6,89.47) (7,94.74) (8,36.84)};
    
    \nextgroupplot[ylabel={Score}, title={Javascript}]
    \addplot [fill=mygreen] coordinates {(0,100)};
    \addplot [fill=mygray] coordinates {(1,100) (2,79.62) (3,69.23) (4,76.92)}; 
    \addplot [fill=myred] coordinates {(5,69.23) (6,46.15) (7,38.46) (8,30.77)};
    
    \nextgroupplot[title={Others}]
    \addplot [fill=mygreen] coordinates {(0,80.95)};
    \addplot [fill=mygray] coordinates {(1,76.19) (2,57.14) (3,47.62) (4,71.43)}; 
    \addplot [fill=myred] coordinates {(5,38.1) (6,61.9) (7,85.71) (8,66.67)};
    
    \nextgroupplot[title={Overall Accuracy}]
    \addplot [fill=mygreen] coordinates {(0,94.86)};
    \addplot [fill=mygray] coordinates {(1,75.42) (2,67.16) (3,60.84) (4,70.34)}; 
    \addplot [fill=myred] coordinates {(5,62.07) (6,63.67) (7,68.15) (8,55.98)};
    
    \end{groupplot}
  \end{tikzpicture}
    \captionof{figure}{Performance comparison of different tools across the overall category and five specific subcategories. The abbreviations used for baseline methods are listed in Table~\ref{table:baseline_tools}.}
  \label{fig:acc-compare-program-language}
\end{figure*}
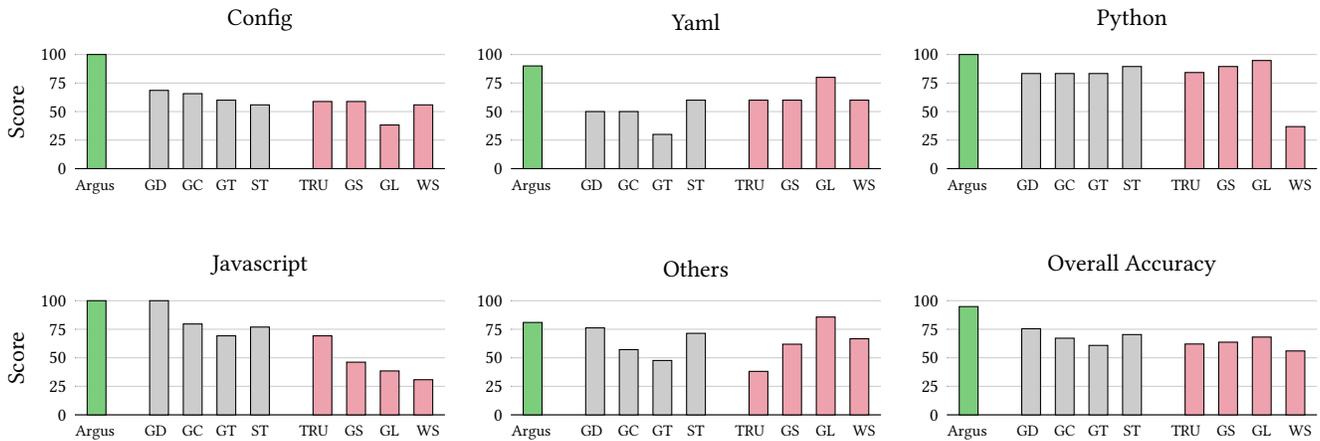

According to Figure~\ref{fig:acc-compare-program-language}, the overall accuracy of Argus is far superior to other methods. Notably, the detection accuracy for \textbf{Config} and YAML files reaches 100\% and 90\%, which is attributed to the fact that Argus does not solely rely on the secret content for its judgment; its semantic analysis of reference relationships and contextual environments allows for a more accurate and comprehensive evaluation of secret leakage. For projects written in Python and JavaScript, the detection accuracy is 100.00\% .

For the \textbf{Config} and \textbf{Other Languages} categories, the variety of types and complexity of analysis conditions are generally more challenging. The detailed detection results for these categories are shown in Table~\ref{tab:config-result-table} and Table~\ref{tab:other-langueges-result-table}, respectively, demonstrating that Argus maintains excellent detection capabilities and performance even in diverse and complex scenarios.

\begin{table}[t]

\centering
\caption{Configuration file detection results}
\label{tab:config-result-table}
\footnotesize

\begin{tabular}{@{}lccccc|cccc@{}}
\toprule
\multirow{2}{*}{Category} & \multicolumn{5}{c|}{AI-driven} & \multicolumn{4}{c}{Regex-driven} \\
& Argus & GD & GC & GT & ST & TRU & GS & GL & WS \\

\midrule
Ipynb    & \good{\textbf{3/3}} & \average{2/3}  & \good{3/3} & \average{2/3} & \average{2/3} & \average{2/3} & \good{3/3} & \good{3/3} & \poor{1/3} \\
Markdown & \good{\textbf{3/3}} & \poor{1/3} & \poor{1/3} & \poor{0/3} & \poor{1/3} & \poor{0/3} & \average{2/3} & \average{2/3} & \good{3/3} \\
Key      & \good{\textbf{3/3}} & \poor{0/3}  & \poor{0/3} & \poor{0/3} & \poor{0/3} & \poor{0/3} & \poor{0/3} & \poor{0/3} & \good{3/3} \\
Pem      & \good{\textbf{2/2}} & \good{2/2} & \good{2/2} & \good{2/2} & \poor{1/2} & \good{2/2} & \good{2/2} & \poor{0/2} & \poor{0/2} \\
Txt      & \good{\textbf{1/1}} & \poor{0/1} & \poor{0/1} & \poor{0/1} & \poor{0/2} & \poor{0/1} & \poor{0/1} & \poor{0/1} & \average{1/2} \\
Conf     & \good{\textbf{1/1}} & \good{1/1} & \poor{0/1} & \poor{0/1} & \good{1/1} & \poor{0/1} & \good{1/1} & \good{1/1} & \good{1/1} \\
Env      & \good{\textbf{9/9}} & \good{7/9} & \good{7/9} & \good{7/9} & \average{5/9} & \good{7/9} & \average{3/9} & \average{3/9} & \average{3/9} \\
Properties & \good{\textbf{4/4}} & \good{4/4} & \good{4/4} & \good{4/4} & \poor{0/4} & \good{4/4} & \poor{0/4} & \poor{0/4} & \poor{0/4} \\
Json     & \good{\textbf{5/5}} & \average{3/5} & \average{3/5} & \average{3/5} & \good{5/5} & \average{3/5} & \good{5/5} & \average{3/5} & \good{4/5} \\
Git      & \good{\textbf{2/2}} & \poor{0/2} & \poor{0/2} & \poor{0/2} & \good{2/2} & \poor{0/2} & \good{2/2} & \good{2/2} & \good{2/2} \\
Data     & \good{\textbf{2/2}} & \good{2/2} & \good{2/2} & \good{2/2} & \good{2/2} & \good{2/2} & \good{2/2} & \poor{0/2} & \poor{0/2} \\
\bottomrule
\end{tabular}

\vspace{0.2cm}
\footnotesize
\begin{minipage}{\linewidth}

\footnotesize\textcolor{gray}{Note: \textcolor{mygreen}{Green} indicates excellent performance; \textcolor{myred}{Red} indicates poor performance; \\
\textcolor{mygray}{Gray} indicates average performance.}
\end{minipage}
\end{table}

\begin{table}[t] 
\centering
\caption{Other Languages Detection Results}
\label{tab:other-langueges-result-table}
\footnotesize

\begin{tabular}{@{}lccccc|cccc@{}}
\toprule
\multirow{2}{*}{Category} & \multicolumn{5}{c|}{AI-driven} & \multicolumn{4}{c}{Regex-driven} \\
& Argus & GD & GC & GT & ST & TRU & GS & GL & WS \\

\midrule
Cs          & \good{\textbf{2/2}} & \good{2/2} & \good{2/2} & \average{1/2} & \average{1/2} & \good{2/2} & \good{2/2} & \good{2/2} & \na{NA}  \\
C           & \good{\textbf{2/2}} & \average{1/2} & \average{1/2} & \poor{0/2} & \good{2/2} & \poor{0/2} & \poor{0/2} & \good{2/2} & \na{NA}  \\
Cpp         & \good{\textbf{1/1}} & \good{1/1} & \poor{0/1} & \good{1/1} & \good{1/1} & \poor{0/1} & \good{1/1} & \good{1/1} & \na{NA}  \\
Dockerfile  & \average{\textbf{1/2}} & \good{2/2} & \good{2/2} & \good{2/2} & \average{1/2} & \good{2/2} & \good{2/2} & \good{2/2} & \average{1/2}  \\
Shell       & \average{\textbf{1/2}} & \average{1/2} & \average{1/2} & \average{1/2} & \good{2/2} & \average{1/2} & \good{2/2} & \good{2/2} & \good{2/2}  \\
Dsl         & \good{\textbf{1/1}} & \poor{0/1} & \poor{0/1} & \poor{0/1} & \good{1/1} & \poor{0/1} & \good{1/1} & \good{1/1} & \na{NA}  \\
Java        & \average{\textbf{3/5}} & \good{3/5} & \average{2/5} & \average{2/5} & \good{3/5} & \average{2/5} & \good{3/5} & \good{3/5} & \good{3/5}  \\
Html        & \good{\textbf{1/1}} & \good{1/1} & \good{1/1} & \good{1/1} & \good{1/1} & \good{1/1} & \good{1/1} & \poor{0/1} & \good{1/1}  \\
Typescript  & \good{\textbf{2/2}} & \good{2/2} & \good{2/2} & \good{2/2} & \good{2/2} & \good{2/2} & \good{2/2} & \poor{0/2} & \good{2/2}  \\
Php         & \good{\textbf{2/2}} & \average{1/2} & \poor{0/2} & \poor{0/2} & \average{1/2} & \poor{0/2} & \poor{0/2} & \average{1/2} & \good{2/2}  \\
Tcl         & \good{\textbf{1/1}} & \good{1/1} & \good{1/1} & \poor{0/1} & \poor{0/1} & \poor{0/1} & \good{1/1} & \na{NA} & \poor{0/1}  \\
\bottomrule
\end{tabular}
\vspace{0.2cm}
\footnotesize

\begin{minipage}{\linewidth}

\footnotesize\textcolor{gray}{Note: \textcolor{mygreen}{Green} indicates excellent performance; \textcolor{myred}{Red} indicates poor performance; \\
\textcolor{mygray}{Gray} indicates average performance; \texttt{NA} indicates that the tool does not support this task.}
\end{minipage}
\end{table}

\subsubsection{Comparison results of secret category detection}

\begin{figure*}
   \centering
   \includegraphics[width=0.8\textwidth]{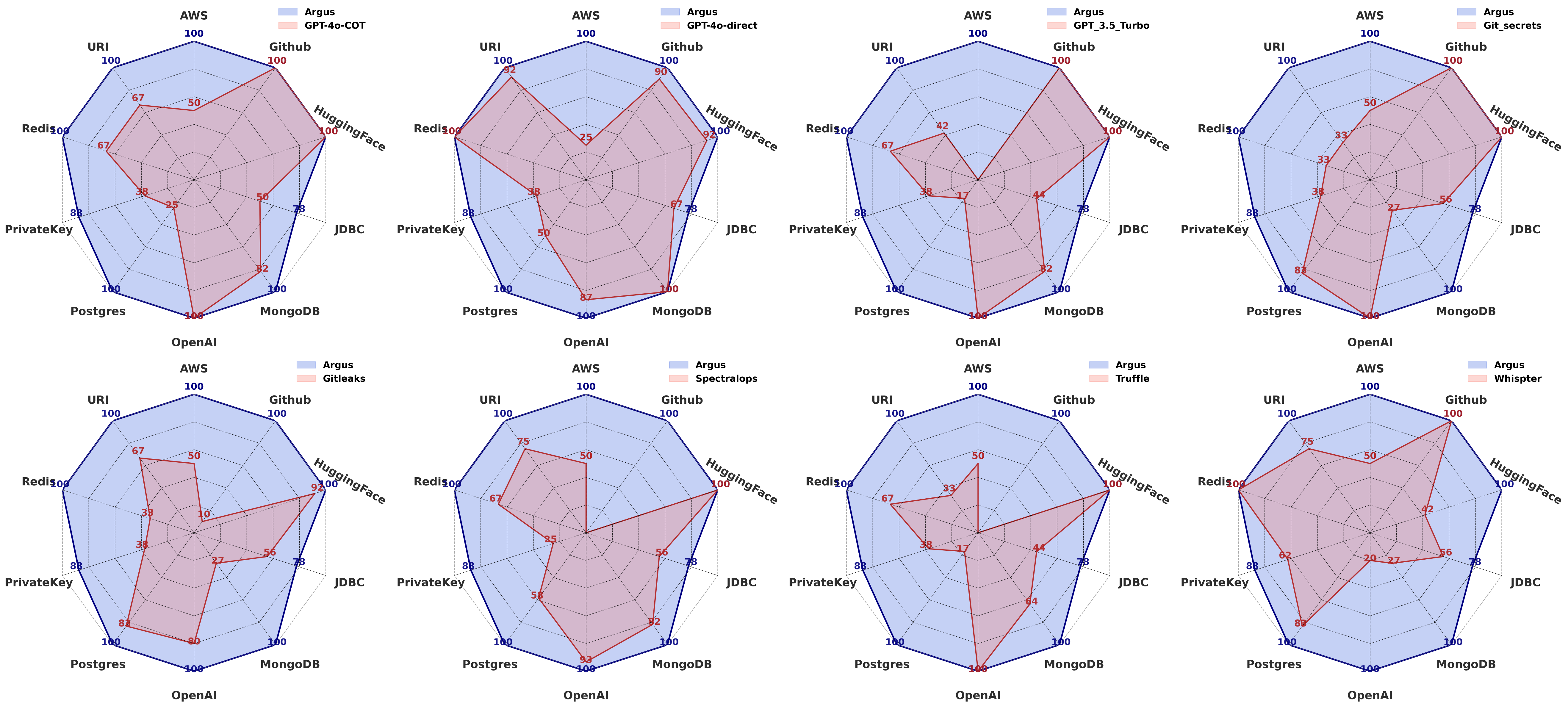}
   \caption{Performance Comparison of Argus and Baseline Methods for Classification by Secret Types}
   \label{fig:Performance_Comparison}
\end{figure*}

Argus outperforms the existing LLM solutions—including GPT-4, GPT-4-CoT, and GPT-3.5—in overall key leakage detection. Experimental results show that Argus achieves a 100\% detection rate for critical key types such as AWS, GitHub, Postgres, and Redis, whereas GPT-4 and its variants exhibit significant bias across several key categories. For example, GPT-4 achieves only a 37.5\% detection rate for \texttt{PrivateKey}, whereas our method attains 87.5\%. Moreover, GPT-4-CoT demonstrates limited detection capability for sensitive keys such as \texttt{Postgres} and \texttt{PrivateKey}—with detection rates even falling below 40\%—possibly due to an over-analysis of key information by the CoT approach affecting the final decision. In contrast, Argus maintains high stability across different key types, avoiding the detection accuracy fluctuations observed in the LLM method, and exhibits stronger generalization and applicability.

Compared with traditional key detection tools (such as Truffle, Git-Secrets, Gitleaks, Whispter, and SpectralOps), Argus demonstrates a more comprehensive detection capability. Experimental results indicate that while traditional tools may perform better on specific key types, they suffer from significant limitations in generalization. For instance, Truffle's detection capability for GitHub keys is nearly 0\%, because data that meets high entropy characteristics is directly classified as a false positive, without distinguishing whether the secret is genuinely leaked. Git-Secrets shows detection rates below 30\% for MongoDB and URI, and Gitleaks only achieves a 10\% detection rate for GitHub keys. Furthermore, some tools (such as SpectralOps) achieve only a 25\% detection rate for \texttt{PrivateKey}, posing a high risk of false negatives. In contrast, our method maintains high detection rates across all key types, with particularly stable performance on core key categories, effectively reducing both false negatives and false positives, and providing a more reliable solution for key leakage detection.

\subsubsection{Overall Performance Evaluation and Ablation Study}

To further validate the superiority of Argus, we adopt TruffleHog’s initial screening results as our baseline and compute Precision, Recall and F1 score for Argus and seven other methods (excluding TruffleHog itself) on the CommonLeak dataset, as shown in Table \ref{tab:precision_recall_F1_score}.  The GPT-4o model with the Chain-of-Thought (CoT) approach achieves a Recall of 100
\% but only a Precision of 63.64\%, resulting in a high false-positive rate.  By contrast, the traditional detection tool WhispeR attains perfect Precision (100\%) yet suffers severe false negatives with a Recall of only 19.64\%.  Argus, however, balances both metrics—achieving 96.36\% Precision and 94.64\% Recall—and attains an F1 score of 0.955, clearly outperforming all competitors.

Moreover, Argus’s detection workflow is organized into three hierarchical levels. To quantify the contribution of each level, we conduct a progressive ablation study. Table \ref{tab:precision_recall_F1_score} reports the impact of removing successive levels on Accuracy, Precision and Recall. With only Level 1 active, detection on CommonLeak yields 76.29\% Accuracy, 79.57\% Precision and 94.87\% Recall.  Adding Level 2  slightly raises Accuracy to 77.32\% and Precision to 83.33\%, while Recall dips modestly to 91.46\%, resulting in only minor overall change.  Finally, when all three levels are integrated, Argus achieves its best performance on CommonLeak: 94.86\% Accuracy, 96.36\% Precision and 94.64\% Recall, with each metric significantly improved over the ablated variants.

\begin{table*}[htbp]
\centering
\caption{Performance Comparison of Argus and Baseline Methods on the CommonLeak Dataset in Terms of Precision, Recall, and F1 Score}
\label{tab:precision_recall_F1_score}
\renewcommand\arraystretch{1.5} 

\scalebox{0.65}{

\begin{tabular}{
    c|
    >{\centering\arraybackslash}p{1.1cm}
    >{\centering\arraybackslash}p{1.1cm}
    >{\centering\arraybackslash}p{1.4cm}
    |
    >{\centering\arraybackslash}p{1.1cm}
    >{\centering\arraybackslash}p{1.1cm}
    >{\centering\arraybackslash}p{1.1cm}
    >{\centering\arraybackslash}p{1.1cm}
    |
    >{\centering\arraybackslash}p{1.1cm}
    >{\centering\arraybackslash}p{1.1cm}
    >{\centering\arraybackslash}p{1.1cm}
    }
\toprule
\multirow{2}{*}{\diagbox{Metric}{Method}} & \multicolumn{3}{c|}{M-A}
  & \multicolumn{4}{c|}{AI-driven} 
  & \multicolumn{3}{c}{Regex-driven} \\
\cmidrule(lr){2-4}\cmidrule(lr){5-8} \cmidrule(lr){9-11}
& Argus & Level 1 & Level 1+2 & ST & GD & GC & GT
& GL & WS & GS \\
\midrule
Accuracy  & \textbf{\myhighlight{94.86}} &76.29 & 77.32 & 74.58 & 69.07 & 75.26 & 67.01 & 59.79 & 53.61 & 63.92 \\
Precision & \myhighlight{96.36} & 79.57 & 83.33 & 70.97 & 72.22 & 63.64 & 59.34 & 78.79 & \textbf{100} & 70.59 \\
Recall    & \myhighlight{94.64} &94.87 & 91.46 & 78.57 & 92.86 & \textbf{100} & 96.43 & 46.43 & 19.64 & 64.29 \\
F1 Score  & \textbf{\myhighlight{95.5}} &86.55 &87.21& 74.58 & 81.25 & 77.78 & 73.47 & 58.43 & 32.85 & 67.29 \\

\bottomrule
\end{tabular}
}
\vspace{0.2cm}
\begin{minipage}{\textwidth}

\footnotesize\textcolor{gray}{Note: M-A stands for Multi-Agent.A denotes Argus with only the first detection level active (Level 1). A + B denotes Argus with both the first and second detection levels active (Levels 1 + 2).}
\end{minipage}
\end{table*}

\subsubsection{Stability Analysis of Multi-Agent System}
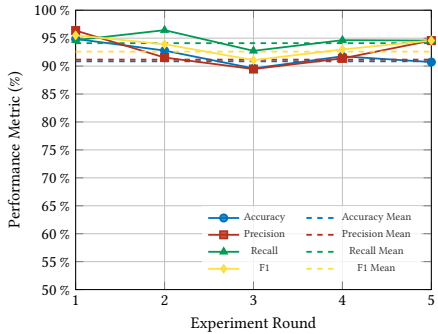
\begin{figure}[htbp]
  \centering
\resizebox{0.7\linewidth}{!}{
  \begin{tikzpicture}
    \begin{axis}[
      width=\linewidth,
      height=7cm,
      xlabel={Experiment Round},
      ylabel={Performance Metric (\%)},
      xmin=1, xmax=5,
      ymin=50, ymax=100,
      xtick={1,2,3,4,5},
      ytick={50,55,60,65,70,75,80,85,90,95,100},
      yticklabel={\pgfmathprintnumber{\tick}\,\%},
      grid=major,
      legend pos=south east,
      legend style={
        font=\scriptsize,
        draw=none,
        fill=white,
        fill opacity=0.8,
        /tikz/column 2/.style={column sep=1em}
      },
      legend columns=2,
      every axis plot/.append style={
        line width=1pt,
        mark size=2pt
      }
    ]

      \addplot[RoyalBlue, mark=*]   coordinates {(1,94.86) (2,92.78) (3,89.58) (4,91.75) (5,90.72)};
      \addlegendentry{Accuracy}
      \addplot[dashed,RoyalBlue]     coordinates {(1,90.85) (5,90.85)};
      \addlegendentry{Accuracy Mean}

      \addplot[BrickRed, mark=square*]   coordinates {(1,96.36) (2,91.53) (3,89.47) (4,91.38) (5,94.57)};
      \addlegendentry{Precision}
      \addplot[dashed,BrickRed]           coordinates {(1,91.16) (5,91.16)};
      \addlegendentry{Precision Mean}

      \addplot[ForestGreen, mark=triangle*]   coordinates {(1,94.64) (2,96.43) (3,92.73) (4,94.64) (5,94.57)};
      \addlegendentry{Recall}
      \addplot[dashed,ForestGreen]             coordinates {(1,94.10) (5,94.10)};
      \addlegendentry{Recall Mean}

      \addplot[Goldenrod, mark=diamond*]   coordinates {(1,95.50) (2,93.91) (3,91.07) (4,92.98) (5,94.57)};
      \addlegendentry{F1}
      \addplot[dashed,Goldenrod]           coordinates {(1,92.59) (5,92.59)};
      \addlegendentry{F1 Mean}

    \end{axis}
  \end{tikzpicture}
  }
  \caption{Stability analysis of multi-round experiments. Solid lines show each round’s metric values; dashed lines indicate the means.}
  \label{fig:multi-round-stability}
\end{figure}

To assess the robustness and stability of the proposed multi-agent system under identical experimental conditions, we conducted five independent rounds of evaluation on the CommonLeak dataset, recording for each round the key performance metrics—accuracy, precision, recall, and F1 score. As shown in Figure \ref{fig:multi-round-stability}, solid lines depict the metric values for each round, while dashed lines indicate their corresponding averages. Across all five trials, the system maintained a consistently high level of stability, with only minimal fluctuations in its performance measures. Although the inherent nondeterminism of large language models and the risk of hallucinations in multi-agent interactions present potential challenges, Argus’s carefully designed architecture effectively suppresses these sources of instability, thereby delivering reliable and robust performance even in dynamic or noisy environments.

\subsection{Analysis of False Positive Retrieval Capability}

Effectively detecting secret leaks is a fundamental capability; however, the ability to identify and filter false positives is equally critical. To provide a more comprehensive evaluation of tools in handling false positives, we conducted additional assessments using the TrustedFalseSecrets dataset. As an extension of Table~\ref{tab:famouse_check}, this analysis focuses specifically on measuring the tools’ effectiveness in distinguishing false positives from actual leaks. To this end, we selected 20 representative false secrets from ten well-known repositories. Furthermore, we included only baseline methods with demonstrated analytical capabilities for false positive detection, as other methods lack the ability to differentiate between false and genuine secrets.

The detection results for these 20 false secrets are presented in Table~\ref{tab:famouse_check}. The results show that the multi-agent system correctly classified all 20 secrets, while the SpectralOps tool correctly identified 12 of them. In comparison, the GPT-4o model using the CoT approach and the standalone GPT-4o model each correctly identified only 11 secrets, and the GPT-3.5 model correctly identified only 7 secrets. These findings demonstrate that Argus, through its multi-agent architecture and three-tier logical analysis, significantly enhances the ability to assess the authenticity of keys.

\begin{table}[t] 
\centering
\caption{Typical secret test results of well-known warehouses}
\label{tab:famouse_check}
\footnotesize
\scalebox{0.9}{
\begin{tabular}{@{}llccccc@{}}
\toprule
Repo & File Name & Argus & GD & GC & GT & SP \\
\midrule
\multirow{3}{*}{moby} 
  & \href{https://github.com/moby/moby/blob/master/integration/testdata/https/server-key.pem}{server-key.pem \,{\scriptsize\faExternalLink}}  
  & \textcolor{checkmarkgreen}{$\checkmark$} & \textcolor{crossred}{$\times$} & \textcolor{crossred}{$\times$} & \textcolor{crossred}{$\times$} & \textcolor{crossred}{$\times$} \\[2mm]
  & \href{https://github.com/moby/moby/blob/master/integration-cli/fixtures/https/client-rogue-key.pem}{client-rogue-key.pem \,{\scriptsize\faExternalLink}}  
  & \textcolor{checkmarkgreen}{$\checkmark$} & \textcolor{crossred}{$\times$} & \textcolor{crossred}{$\times$} & \textcolor{crossred}{$\times$} & \textcolor{crossred}{$\times$} \\[2mm]
  & \href{https://github.com/moby/moby/blob/master/registry/config_test.go\#L38}{config\_test.go \,{\scriptsize\faExternalLink}}  
  & \textcolor{checkmarkgreen}{$\checkmark$} & \textcolor{checkmarkgreen}{$\checkmark$} & \textcolor{checkmarkgreen}{$\checkmark$} & \textcolor{checkmarkgreen}{$\checkmark$} & \textcolor{checkmarkgreen}{$\checkmark$} \\
\midrule

\multirow{2}{*}{kubernetes}  
  & \href{https://github.com/kubernetes/kubernetes/blob/master/staging/src/k8s.io/kubectl/pkg/cmd/create/create_secret_tls_test.go\#L54}{create\_secret\_tls\_test.go \,{\scriptsize\faExternalLink}}  
  & \textcolor{checkmarkgreen}{$\checkmark$} & \textcolor{crossred}{$\times$} & \textcolor{crossred}{$\times$} & \textcolor{crossred}{$\times$} & \textcolor{crossred}{$\times$} \\[2mm]
  & \href{https://github.com/kubernetes/kubernetes/blob/master/test/cmd/kubeconfig.sh\#L152}{kubeconfig.sh \,{\scriptsize\faExternalLink}}  
  & \textcolor{checkmarkgreen}{$\checkmark$} & \textcolor{crossred}{$\times$} & \textcolor{crossred}{$\times$} & \textcolor{checkmarkgreen}{$\checkmark$} & \textcolor{crossred}{$\times$} \\
\midrule

\multirow{2}{*}{bitcoin}  
  & \href{https://github.com/bitcoin/bitcoin/blob/master/src/validation.cpp\#L13}{validation.cpp \,{\scriptsize\faExternalLink}}  
  & \textcolor{checkmarkgreen}{$\checkmark$} & \textcolor{checkmarkgreen}{$\checkmark$} & \textcolor{checkmarkgreen}{$\checkmark$} & \textcolor{checkmarkgreen}{$\checkmark$} & \textcolor{checkmarkgreen}{$\checkmark$} \\[2mm]
  & \href{https://github.com/bitcoin/bitcoin/blob/master/test/functional/feature_taproot.py\#L1537}{feature\_taproot.py \,{\scriptsize\faExternalLink}}  
  & \textcolor{checkmarkgreen}{$\checkmark$} & \textcolor{checkmarkgreen}{$\checkmark$} & \textcolor{checkmarkgreen}{$\checkmark$} & \textcolor{checkmarkgreen}{$\checkmark$} & \textcolor{checkmarkgreen}{$\checkmark$} \\
\midrule

\multirow{2}{*}{webpack}  
  & \href{https://github.com/webpack/webpack/blob/main/test/hotCases/lazy-compilation/https/key.pem}{key.pem \,{\scriptsize\faExternalLink}}  
  & \textcolor{checkmarkgreen}{$\checkmark$} & \textcolor{crossred}{$\times$}  & \textcolor{crossred}{$\times$}  & \textcolor{crossred}{$\times$}  & \textcolor{crossred}{$\times$} \\[2mm]
  & \href{https://github.com/webpack/webpack/blob/main/test/SharingUtil.unittest.js\#L165}{SharingUtil.unittest.js \,{\scriptsize\faExternalLink}}  
  & \textcolor{checkmarkgreen}{$\checkmark$} & \textcolor{crossred}{$\times$}  & \textcolor{crossred}{$\times$}  & \textcolor{crossred}{$\times$}  & \textcolor{checkmarkgreen}{$\checkmark$} \\
\midrule

\multirow{2}{*}{spring-boot}  
  & \href{https://github.com/spring-projects/spring-boot/blob/main/spring-boot-project/spring-boot-actuator-autoconfigure/src/test/java/org/springframework/boot/actuate/autoconfigure/cloudfoundry/reactive/ReactiveTokenValidatorTests.java\#L273}{ReactiveTokenValidatorTests.java \,{\scriptsize\faExternalLink}}  
  & \textcolor{checkmarkgreen}{$\checkmark$} & \textcolor{crossred}{$\times$} & \textcolor{crossred}{$\times$} & \textcolor{crossred}{$\times$} & \textcolor{crossred}{$\times$} \\[2mm]
  & \href{https://github.com/spring-projects/spring-boot/blob/main/spring-boot-project/spring-boot-autoconfigure/src/dockerTest/resources/org/springframework/boot/autoconfigure/mail/ssl/test-server.key}{test-server.key \,{\scriptsize\faExternalLink}}  
  & \textcolor{checkmarkgreen}{$\checkmark$} & \textcolor{crossred}{$\times$} & \textcolor{crossred}{$\times$} & \textcolor{crossred}{$\times$} & \textcolor{crossred}{$\times$} \\
\midrule

\multirow{1}{*}{fastapi}  
  & \href{https://github.com/fastapi/fastapi/blob/master/docs/en/data/people.yml\#L1213}{people.yml \,{\scriptsize\faExternalLink}}  
  & \textcolor{checkmarkgreen}{$\checkmark$} & \textcolor{checkmarkgreen}{$\checkmark$} & \textcolor{crossred}{$\times$}  & \textcolor{crossred}{$\times$}  & \textcolor{checkmarkgreen}{$\checkmark$} \\
\midrule

\multirow{2}{*}{neovim} 
  & \href{https://github.com/neovim/neovim}{.git/pack-954a***8e3b.pack \,{\scriptsize\faExternalLink}}  
  & \textcolor{checkmarkgreen}{$\checkmark$} & \textcolor{checkmarkgreen}{$\checkmark$} & \textcolor{checkmarkgreen}{$\checkmark$} & \textcolor{checkmarkgreen}{$\checkmark$} & \textcolor{checkmarkgreen}{$\checkmark$} \\[2mm]
  & \href{https://github.com/neovim/neovim/blob/master/src/gen/gen_help_html.lua\#L1002}{gen\_help\_html.lua \,{\scriptsize\faExternalLink}}  
  & \textcolor{checkmarkgreen}{$\checkmark$} & \textcolor{checkmarkgreen}{$\checkmark$} & \textcolor{checkmarkgreen}{$\checkmark$} & \textcolor{checkmarkgreen}{$\checkmark$} & \textcolor{checkmarkgreen}{$\checkmark$} \\
\midrule

\multirow{2}{*}{pandas}  
  & \href{https://github.com/pandas-dev/pandas/blob/main/pandas/tests/io/formats/style/test_html.py\#L787}{test\_html.py \,{\scriptsize\faExternalLink}}  
  & \textcolor{checkmarkgreen}{$\checkmark$} & \textcolor{crossred}{$\times$} & \textcolor{checkmarkgreen}{$\checkmark$} & \textcolor{crossred}{$\times$} & \textcolor{crossred}{$\times$} \\[2mm]
  & \href{https://github.com/pandas-dev/pandas/blob/main/pandas/tests/io/formats/style/test_html.py\#L790}{test\_html.py \,{\scriptsize\faExternalLink}}  
  & \textcolor{checkmarkgreen}{$\checkmark$} & \textcolor{checkmarkgreen}{$\checkmark$} & \textcolor{checkmarkgreen}{$\checkmark$} & \textcolor{checkmarkgreen}{$\checkmark$} & \textcolor{checkmarkgreen}{$\checkmark$} \\
\midrule

\multirow{2}{*}{vue}  
  & \href{https://github.com/vuejs/vue/blob/main/CHANGELOG.md\#L718}{CHANGELOG.md \,{\scriptsize\faExternalLink}}  
  & \textcolor{checkmarkgreen}{$\checkmark$} & \textcolor{checkmarkgreen}{$\checkmark$} & \textcolor{checkmarkgreen}{$\checkmark$} & \textcolor{crossred}{$\times$}  & \textcolor{checkmarkgreen}{$\checkmark$} \\[2mm]
  & \href{https://github.com/vuejs/vue/blob/main/CHANGELOG.md\#L813}{CHANGELOG.md \,{\scriptsize\faExternalLink}}  
  & \textcolor{checkmarkgreen}{$\checkmark$} & \textcolor{checkmarkgreen}{$\checkmark$} & \textcolor{checkmarkgreen}{$\checkmark$} & \textcolor{crossred}{$\times$}  & \textcolor{checkmarkgreen}{$\checkmark$} \\
\midrule

\multirow{2}{*}{transformers} 
  & \href{https://github.com/huggingface/transformers}{.git/packed-refs \,{\scriptsize\faExternalLink}}  
  & \textcolor{checkmarkgreen}{$\checkmark$} & \textcolor{checkmarkgreen}{$\checkmark$} & \textcolor{checkmarkgreen}{$\checkmark$} & \textcolor{crossred}{$\times$}  & \textcolor{checkmarkgreen}{$\checkmark$} \\[2mm]
  & \href{https://github.com/huggingface/transformers/blob/main/src/transformers/testing_utils.py\#L195}{testing\_utils.py \,{\scriptsize\faExternalLink}}  
  & \textcolor{checkmarkgreen}{$\checkmark$} & \textcolor{checkmarkgreen}{$\checkmark$} & \textcolor{checkmarkgreen}{$\checkmark$} & \textcolor{crossred}{$\times$}  & \textcolor{checkmarkgreen}{$\checkmark$} \\
\bottomrule
\end{tabular}
}

\begin{minipage}{\linewidth}
\footnotesize\textcolor{gray}{Note: \textcolor{checkmarkgreen}{$\checkmark$} indicates a correct detection result, while \textcolor{crossred}{$\times$} represents an incorrect detection result. Due to the limitations of traditional detection methods, the typical false positive cases we selected could not be identified, making the detection accuracy appear relatively high. However, this does not accurately reflect the actual performance, and thus these cases were not included in the table.}
\end{minipage}
\end{table}

\subsection{Cost Analysis}
\begin{table}[h]
    \caption{\centering Comparison of Cost, Efficiency, and Accuracy Across Various LLMs}
    \centering
    \renewcommand{\arraystretch}{1.3}  
    \resizebox{0.9\linewidth}{!}{  
    \begin{tabular}{lccccc}
        \toprule
        \multirow{2}{*}{Method} & Cost & Avg. Cost & Time & Avg. Time & Accuracy \\
        & (Dollar) & (Dollar) & (Minute) & (Minute) & (\%) \\
        \midrule
        Argus        & 2.21  & 0.023 & 68.00 & 0.70 & 94.86 \\
        GPT-4-direct & 0.42  & 0.004 & 9.00  & 0.09 & 75.26 \\
        GPT-4-COT    & 0.57  & 0.006 & 19.68 & 0.20 & 67.01 \\
        GPT-3.5      & 0.087 & 0.001 & 10.66 & 0.11 & 59.79 \\
        \bottomrule
    \end{tabular}
    }
    \label{tab:cost_analyze}
\end{table}

As the first sensitive‐information detection tool driven by a large language model, Argus has no directly comparable baseline in terms of LLM token consumption, and the resource profiles of traditional scanners are not commensurable. Accordingly, we benchmarked Argus against a GPT-4–based direct detection approach to compare cost and runtime. As Table \ref{tab:cost_analyze} shows, although Argus relies on a multi‐agent system to perform extensive relational analysis, its use of an information pool for inter‐agent data sharing and node‐level summary compression substantially reduces redundant computation. On the CommonLeak dataset—comprising 97 real-world code repositories—Argus incurs a total detection cost of just \$2.21 and requires only 68 minutes for single-threaded execution, demonstrating excellent computational efficiency. While the multi‐agent architecture inevitably introduces some overhead through additional reasoning and iterations, the marked improvement in detection accuracy renders this cost entirely acceptable.

\section{RELATED WORK}

\label{sec:rel_work}

\textbf{Traditional Secret Detection Methods.}Existing open-source secret leakage detection tools primarily use regular expression matching or entropy calculation to identify potential leaked strings \cite{10063545}. Their focuses differ: for example, TruffleHog and Gitleaks employ a complete rule set to screen for all leaks, while Git-Secrets uses hooks and user-specified regex sets to enhance flexibility. However, these tools cannot analyze whether a secret itself contains readable placeholders, nor can they understand the context in which a secret appears or whether it is referenced by other files within the project \cite{10292698}. As a result, the scan results contain a large number of false positives \cite{Smith2020WhyCJ} \cite{10.1145/3510003.3510214}. Consequently, more researchers have turned their attention to a screening model that combines regular expressions with machine learning \cite{rahman2022secret}. However, such methods share common issues. First, using confidential information to train the model carries the risk of leaking secrets. Second, since the dataset involves confidential information, it cannot be shared. Runhan Feng et al. built a deep neural network for automated detection and achieved certain results. However, their approach still suffers from limitations such as low scalability and limited coverage \cite{feng2022automated}.

\textbf{Multi-Agent System.} With the enhancement of LLMs in semantic understanding and reasoning capabilities, LLM-based autonomous agents have gradually emerged. Li Wang et al. \cite{wang2024survey} proposed a framework for constructing LLM agents, which primarily consists of modules such as role definition, memory, planning, and action. Theodore R. Sumers et al. \cite{sumers2023cognitive} explored the classification of agent memory, categorizing it into working memory and long-term memory, and defined two types of action spaces: external and internal. The external action space involves environmental interaction and communication, while the internal action space includes retrieval, reasoning, and learning. Xiaofei Dong et al. \cite{dong2024survey} conducted a comprehensive survey on LLM-based agents from four perspectives: theoretical foundations, key technologies, application scenarios, and development recommendations.

LLM-based agents have been applied to code generation \cite{zhang2024codeagent,koziolek2024llm} and educational assistance \cite{jin2024teach,li2024edumas}. Research has since advanced to cover agent communication patterns \cite{guo2024large}, automated code-generation methods and their challenges \cite{ramirez2024transforming}, as well as broader open issues \cite{xi2025rise}. These multi-agent systems now underpin applications in software engineering \cite{hong2023metagpt,cinkusz2024towards,wang2025reflexgen}, product design \cite{ding2023designgpt}, digital-twin simulation \cite{xia2024llm}, financial markets \cite{li2025hedgeagents}, embodied robotics \cite{kannan2024smart}, autonomous driving \cite{jiang2024koma}, and code review \cite{islam2024mapcoder,jin2024rgd,wang2024unity,RAGen2025}, demonstrating their wide-ranging potential.However, there is currently no precedent for applying LLM-based agent technology to the field of secret detection. Therefore, we aim to leverage the planning, reasoning, and action capabilities of intelligent agents to advance secret detection technology.

\section{DISCUSSION}

\label{sec:discussion}

The Argus framework offers substantial practical value and supports multiple deployment modes. In its most comprehensive configuration, Argus can employ a LLM as the sole analysis engine, conducting an end-to-end scan of the entire repository. Alternatively, to reduce computational overhead, Argus can integrate with existing secret-detection tools—such as TruffleHog—for an initial localization of candidate secrets. These candidates are then fed into Argus’s multi-agent verification pipeline, which filters out false positives and produces a detailed evidentiary report. In this work, we adopt the latter hybrid approach—using TruffleHog for the initial pass—in order to strike a balance between detection coverage and resource efficiency.

Despite these strengths, Argus still faces certain limitations. First, its initial screening agent depends entirely on TruffleHog’s high-entropy heuristics and rule-based pattern matching. Secrets that fall outside of its predefined rules or involve specialized domain contexts may be overlooked, and downstream agents cannot compensate for these blind spots. Second, Argus currently lacks a reliable mechanism for validating secrets that bear no explicit relationship to code references or links—such as private keys or credentials that reside in the repository but are never referenced by any file. Absent contextual clues, the framework cannot accurately assess their authenticity or operational relevance.

\section{CONCLUSION}
\label{sec:conclusion}

In this work, we introduced Argus, a multi-agent framework designed to mitigate the high false-positive rates frequently observed in traditional sensitive-information detection methods. By leveraging a three-tier mechanism—encompassing key content, file context, and project reference relationships—Argus markedly enhances detection accuracy and reduces manual screening efforts.

Our evaluation on two newly proposed benchmarks demonstrates that Argus achieves a 94\% leak detection accuracy and completely filters out false positives, outperforming existing tools. Moreover, applying Argus to 97 real repositories incurred a cost of only \$2.21, illustrating its efficiency and practicality. These findings highlight Argus's potential as a robust solution for sensitive-information detection and pave the way for future research aimed at extending its capabilities to a broader array of sensitive data types.

\begin{acks}
This work is supported by Guangdong Provincial Key Laboratory of Ultra High Definition Immersive Media Technology(Grant No. 2024B1212010006)
\end{acks}

\bibliographystyle{ACM-Reference-Format}
\bibliography{ref}

@article{guo2024large,
  title={Large language model based multi-agents: A survey of progress and challenges},
  author={Guo, Taicheng and Chen, Xiuying and Wang, Yaqi and Chang, Ruidi and Pei, Shichao and Chawla, Nitesh V and Wiest, Olaf and Zhang, Xiangliang},
  journal={arXiv preprint arXiv:2402.01680},
  year={2024}
}

@article{sumers2023cognitive,
  title={Cognitive architectures for language agents},
  author={Sumers, Theodore and Yao, Shunyu and Narasimhan, Karthik and Griffiths, Thomas},
  journal={Transactions on Machine Learning Research},
  year={2023}
}

@inproceedings{wang2025reflexgen,
  title={RefleXGen: The unexamined code is not worth using},
  author={Wang, Bin and Li, Hui and Liu, AoFan and Yang, BoTao and Yang, Ao and Zhong, YiLu and Huang, Weixiang and Huang, Runhuai and Zeng, Weimin and Zhang, Yanping},
  booktitle={ICASSP 2025-2025 IEEE International Conference on Acoustics, Speech and Signal Processing (ICASSP)},
  pages={1--5},
  year={2025},
  organization={IEEE}
}

@INPROCEEDINGS{RAGen2025,
  author={Liu, Aofan and Li, Haoxuan and Wang, Bin and Yang, Ao and Li, Hui},
  booktitle={2025 International Joint Conference on Neural Networks (IJCNN)}, 
  title={RA-Gen: A Controllable Code Generation Framework Using ReAct for Multi-Agent Task Execution}, 
  year={2025},
  volume={},
  number={},
  pages={1-9},
  doi={10.1109/IJCNN64981.2025.11228706}}

@article{chess2004static,
  title={Static analysis for security},
  author={Chess, Brian and McGraw, Gary},
  journal={IEEE security \& privacy},
  volume={2},
  number={6},
  pages={76--79},
  year={2004},
  publisher={IEEE}
}

@misc{feng2022automated,
  title={Automated Detection of Password Leakage from Public GitHub Repositories. In 2022 IEEE/ACM 44th International Conference on Software Engineering (ICSE). 175--186},
  author={Feng, Runhan and Yan, Ziyang and Peng, Shiyan and Zhang, Yuanyuan},
  year={2022}
}

@inproceedings{lounici2021optimizing,
  title={Optimizing Leak Detection in Open-source Platforms with Machine Learning Techniques.},
  author={Lounici, Sofiane and Rosa, Marco and Negri, Carlo Maria and Trabelsi, Slim and {\"O}nen, Melek},
  booktitle={ICISSP},
  pages={145--159},
  year={2021}
}

@article{islam2024mapcoder,
  title={Mapcoder: Multi-agent code generation for competitive problem solving},
  author={Islam, Md Ashraful and Ali, Mohammed Eunus and Parvez, Md Rizwan},
  journal={arXiv preprint arXiv:2405.11403},
  year={2024}
}

@article{zhang2024codeagent,
  title={Codeagent: Enhancing code generation with tool-integrated agent systems for real-world repo-level coding challenges},
  author={Zhang, Kechi and Li, Jia and Li, Ge and Shi, Xianjie and Jin, Zhi},
  journal={arXiv preprint arXiv:2401.07339},
  year={2024}
}

@inproceedings{jin2024teach,
  title={Teach ai how to code: Using large language models as teachable agents for programming education},
  author={Jin, Hyoungwook and Lee, Seonghee and Shin, Hyungyu and Kim, Juho},
  booktitle={Proceedings of the 2024 CHI Conference on Human Factors in Computing Systems},
  pages={1--28},
  year={2024}
}

@article{hong2023metagpt,
  title={Metagpt: Meta programming for multi-agent collaborative framework},
  author={Hong, Sirui and Zheng, Xiawu and Chen, Jonathan and Cheng, Yuheng and Wang, Jinlin and Zhang, Ceyao and Wang, Zili and Yau, Steven Ka Shing and Lin, Zijuan and Zhou, Liyang and others},
  journal={arXiv preprint arXiv:2308.00352},
  volume={3},
  number={4},
  pages={6},
  year={2023}
}

@inproceedings{Ami_2024,
   title={“False negative - that one is going to kill you”: Understanding Industry Perspectives of Static Analysis based Security Testing},
   url={http://dx.doi.org/10.1109/SP54263.2024.00019},
   DOI={10.1109/sp54263.2024.00019},
   booktitle={2024 IEEE Symposium on Security and Privacy (SP)},
   publisher={IEEE},
   author={Ami, Amit Seal and Moran, Kevin and Poshyvanyk, Denys and Nadkarni, Adwait},
   year={2024},
   month=may, pages={3979–3997} }

@ARTICLE{10305541,
  author={Guo, Zhaoqiang and Tan, Tingting and Liu, Shiran and Liu, Xutong and Lai, Wei and Yang, Yibiao and Li, Yanhui and Chen, Lin and Dong, Wei and Zhou, Yuming},
  journal={IEEE Transactions on Software Engineering}, 
  title={Mitigating False Positive Static Analysis Warnings: Progress, Challenges, and Opportunities}, 
  year={2023},
  volume={49},
  number={12},
  pages={5154-5188},
  doi={10.1109/TSE.2023.3329667}}

@misc{trunk_gitleaks_vs_trufflehog,
  author       = {Trunk.io},
  title        = {Gitleaks vs TruffleHog: Comparing Secret Scanning Tools},
  year         = {2024},
  url          = {https://trunk.io/learn/gitleaks-vs-trufflehog-comparing-secret-scanning-tools#comparing-gitleaks-and-trufflehog-features},
  note         = {Accessed: 2024-12-13}
}

@inproceedings{basak2023comparative,
  title={A comparative study of software secrets reporting by secret detection tools},
  author={Basak, Setu Kumar and Cox, Jamison and Reaves, Bradley and Williams, Laurie},
  booktitle={2023 ACM/IEEE International Symposium on Empirical Software Engineering and Measurement (ESEM)},
  pages={1--12},
  year={2023},
  organization={IEEE}
}

@article{rahman2022secret,
  title={Why secret detection tools are not enough: It’s not just about false positives-an industrial case study},
  author={Rahman, Md Rayhanur and Imtiaz, Nasif and Storey, Margaret-Anne and Williams, Laurie},
  journal={Empirical Software Engineering},
  volume={27},
  number={3},
  pages={59},
  year={2022},
  publisher={Springer}
}

@inproceedings{lazarine2020identifying,
  title={Identifying vulnerable GitHub repositories and users in scientific cyberinfrastructure: An unsupervised graph embedding approach},
  author={Lazarine, Ben and Samtani, Sagar and Patton, Mark and Zhu, Hongyi and Ullman, Steven and Ampel, Benjamin and Chen, Hsinchun},
  booktitle={2020 IEEE International Conference on Intelligence and Security Informatics (ISI)},
  pages={1--6},
  year={2020},
  organization={IEEE}
}

@inproceedings{zhang2023don,
  title={Don't leak your keys: Understanding, measuring, and exploiting the appsecret leaks in mini-programs},
  author={Zhang, Yue and Yang, Yuqing and Lin, Zhiqiang},
  booktitle={Proceedings of the 2023 ACM SIGSAC Conference on Computer and Communications Security},
  pages={2411--2425},
  year={2023}
}

@INPROCEEDINGS{10063545,
  author={Wen, Elliott and Wang, Jia and Dietrich, Jens},
  booktitle={2022 IEEE International Conference on Trust, Security and Privacy in Computing and Communications (TrustCom)}, 
  title={SecretHunter: A Large-scale Secret Scanner for Public Git Repositories}, 
  year={2022},
  volume={},
  number={},
  pages={123-130},
  doi={10.1109/TrustCom56396.2022.00028}}

@inproceedings{basak2023secretbench,
  title={Secretbench: A dataset of software secrets},
  author={Basak, Setu Kumar and Neil, Lorenzo and Reaves, Bradley and Williams, Laurie},
  booktitle={2023 IEEE/ACM 20th International Conference on Mining Software Repositories (MSR)},
  pages={347--351},
  year={2023},
  organization={IEEE}
}

@inproceedings{meli2019bad,
  title={How bad can it git? characterizing secret leakage in public github repositories.},
  author={Meli, Michael and McNiece, Matthew R and Reaves, Bradley}
}

@misc{gitguardian2024secrets,
  author       = {GitGuardian},
  title        = {State of Secrets Sprawl Report 2024},
  year         = {2024},
  url          = {https://www.gitguardian.com/state-of-secrets-sprawl-report-2024},
  note         = {Accessed: 2024-12-13}
}

@misc{trufflehog_github,
  author       = {Truffle Security},
  title        = {TruffleHog: Find leaked credentials},
  year         = {2024},
  url          = {https://github.com/trufflesecurity/trufflehog},
  note         = {Accessed: 2024-12-13}
}

@misc{spectralops_website,
  author       = {SpectralOps},
  title        = {SpectralOps: Automated Secrets Detection and Security Scanning},
  year         = {2024},
  url          = {https://get.spectralops.io/},
  note         = {Accessed: 2024-12-13}
}

@misc{whispers_github,
  author       = {Skyscanner},
  title        = {Whispers: a static code analysis tool},
  year         = {2024},
  url          = {https://github.com/Skyscanner/whispers},
  note         = {Accessed: 2024-12-13}
}

@misc{gitleaks_github,
  author       = {gitleaks},
  title        = {Gitleaks: a tool for detecting secrets},
  year         = {2024},
  url          = {https://github.com/gitleaks/gitleaks},
  note         = {Accessed: 2024-12-13}
}

@misc{gitsecrets_github,
  author       = {AWS Labs},
  title        = {Git-Secrets: Prevents Committing Secrets in Git Repositories},
  year         = {2024},
  url          = {https://github.com/awslabs/git-secrets},
  note         = {Accessed: 2024-12-13}
}

@article{li2025hedgeagents,
  title={HedgeAgents: A Balanced-aware Multi-agent Financial Trading System},
  author={Li, Xiangyu and Zeng, Yawen and Xing, Xiaofen and Xu, Jin and Xu, Xiangmin},
  journal={arXiv preprint arXiv:2502.13165},
  year={2025}
}

@inproceedings{saha2020secrets,
  title={Secrets in source code: Reducing false positives using machine learning},
  author={Saha, Aakanksha and Denning, Tamara and Srikumar, Vivek and Kasera, Sneha Kumar},
  booktitle={2020 International Conference on COMmunication Systems \& NETworkS (COMSNETS)},
  pages={168--175},
  year={2020},
  organization={IEEE}
}

@article{wang2024survey,
  title={A survey on large language model based autonomous agents},
  author={Wang, Lei and Ma, Chen and Feng, Xueyang and Zhang, Zeyu and Yang, Hao and Zhang, Jingsen and Chen, Zhiyuan and Tang, Jiakai and Chen, Xu and Lin, Yankai and others},
  journal={Frontiers of Computer Science},
  volume={18},
  number={6},
  pages={186345},
  year={2024},
  publisher={Springer}
}

@inproceedings{cinkusz2024towards,
  title={Towards LLM-augmented multiagent systems for agile software engineering},
  author={Cinkusz, Konrad and Chudziak, Jaroslaw A},
  booktitle={Proceedings of the 39th IEEE/ACM International Conference on Automated Software Engineering},
  pages={2476--2477},
  year={2024}
}

@inproceedings{xia2024llm,
  title={LLM experiments with simulation: Large Language Model Multi-Agent System for Simulation Model Parametrization in Digital Twins},
  author={Xia, Yuchen and Dittler, Daniel and Jazdi, Nasser and Chen, Haonan and Weyrich, Michael},
  booktitle={2024 IEEE 29th International Conference on Emerging Technologies and Factory Automation (ETFA)},
  pages={1--4},
  year={2024},
  organization={IEEE}
}

@inproceedings{li2024edumas,
  title={EduMAS: A Novel LLM-Powered Multi-Agent Framework for Educational Support},
  author={Li, Qiaomu and Xie, Ying and Chakravarty, Sumit and Lee, Dabae},
  booktitle={2024 IEEE International Conference on Big Data (BigData)},
  pages={8309--8316},
  year={2024},
  organization={IEEE}
}

@inproceedings{ramirez2024transforming,
  title={Transforming Software Development: A Study on the Integration of Multi-Agent Systems and Large Language Models for Automatic Code Generation},
  author={Ram{\'\i}rez-Rueda, Rolando and Ben{\'\i}tez-Guerrero, Edgard and Mezura-Godoy, Carmen and B{\'a}rcenas, Everardo},
  booktitle={2024 12th International Conference in Software Engineering Research and Innovation (CONISOFT)},
  pages={11--20},
  year={2024},
  organization={IEEE}
}

@inproceedings{wang2024unity,
  title={Unity Is Strength: Collaborative LLM-Based Agents for Code Reviewer Recommendation},
  author={Wang, Luqiao and Zhou, Yangtao and Zhuang, Huiying and Li, Qingshan and Cui, Di and Zhao, Yutong and Wang, Lu},
  booktitle={Proceedings of the 39th IEEE/ACM International Conference on Automated Software Engineering},
  pages={2235--2239},
  year={2024}
}

@misc{ding2023designgpt,
  title={DesignGPT: Multi-Agent Collaboration in Design. arXiv},
  author={Ding, S and Chen, X and Fang, Y and Liu, W and Qiu, Y and Chai, C},
  year={2023}
}

@inproceedings{dong2024survey,
  title={A Survey of LLM-based Agents: Theories, Technologies, Applications and Suggestions},
  author={Dong, Xiaofei and Zhang, Xueqiang and Bu, Weixin and Zhang, Dan and Cao, Feng},
  booktitle={2024 3rd International Conference on Artificial Intelligence, Internet of Things and Cloud Computing Technology (AIoTC)},
  pages={407--413},
  year={2024},
  organization={IEEE}
}

@inproceedings{kannan2024smart,
  title={Smart-llm: Smart multi-agent robot task planning using large language models},
  author={Kannan, Shyam Sundar and Venkatesh, Vishnunandan LN and Min, Byung-Cheol},
  booktitle={2024 IEEE/RSJ International Conference on Intelligent Robots and Systems (IROS)},
  pages={12140--12147},
  year={2024},
  organization={IEEE}
}

@article{jiang2024koma,
  title={Koma: Knowledge-driven multi-agent framework for autonomous driving with large language models},
  author={Jiang, Kemou and Cai, Xuan and Cui, Zhiyong and Li, Aoyong and Ren, Yilong and Yu, Haiyang and Yang, Hao and Fu, Daocheng and Wen, Licheng and Cai, Pinlong},
  journal={IEEE Transactions on Intelligent Vehicles},
  year={2024},
  publisher={IEEE}
}

@article{xi2025rise,
  title={The rise and potential of large language model based agents: A survey},
  author={Xi, Zhiheng and Chen, Wenxiang and Guo, Xin and He, Wei and Ding, Yiwen and Hong, Boyang and Zhang, Ming and Wang, Junzhe and Jin, Senjie and Zhou, Enyu and others},
  journal={Science China Information Sciences},
  volume={68},
  number={2},
  pages={121101},
  year={2025},
  publisher={Springer}
}

@inproceedings{jin2024rgd,
  title={RGD: Multi-LLM Based Agent Debugger via Refinement and Generation Guidance},
  author={Jin, Haolin and Sun, Zechao and Chen, Huaming},
  booktitle={2024 IEEE International Conference on Agents (ICA)},
  pages={136--141},
  year={2024},
  organization={IEEE}
}

@inproceedings{koziolek2024llm,
  title={Llm-based control code generation using image recognition},
  author={Koziolek, Heiko and Koziolek, Anne},
  booktitle={Proceedings of the 1st International Workshop on Large Language Models for Code},
  pages={38--45},
  year={2024}
}

@article{nunez2024autosafecoder,
  title={Autosafecoder: A multi-agent framework for securing llm code generation through static analysis and fuzz testing},
  author={Nunez, Ana and Islam, Nafis Tanveer and Jha, Sumit Kumar and Najafirad, Peyman},
  journal={arXiv preprint arXiv:2409.10737},
  year={2024}
}

@ARTICLE{10292698,
  author={Han, Ruidong and Gong, Huihui and Ma, Siqi and Li, Juanru and Xu, Chang and Bertino, Elisa and Nepal, Surya and Ma, Zhuo and Ma, Jianfeng},
  journal={IEEE Transactions on Information Forensics and Security}, 
  title={A Credential Usage Study: Flow-Aware Leakage Detection in Open-Source Projects}, 
  year={2024},
  volume={19},
  number={},
  pages={722-734},
  doi={10.1109/TIFS.2023.3326985}}

@inproceedings{Smith2020WhyCJ,
  title={Why Can't Johnny Fix Vulnerabilities: A Usability Evaluation of Static Analysis Tools for Security},
  author={Justin Smith and Lisa Nguyen Quang Do and Emerson R. Murphy-Hill},
  booktitle={SOUPS @ USENIX Security Symposium},
  year={2020},
  url={https://api.semanticscholar.org/CorpusID:220359988}
}

@inproceedings{10.1145/3510003.3510214,
author = {Kang, Hong Jin and Aw, Khai Loong and Lo, David},
title = {Detecting false alarms from automatic static analysis tools: how far are we?},
year = {2022},
isbn = {9781450392211},
publisher = {Association for Computing Machinery},
address = {New York, NY, USA},
url = {https://doi.org/10.1145/3510003.3510214},
doi = {10.1145/3510003.3510214},
booktitle = {Proceedings of the 44th International Conference on Software Engineering},
pages = {698–709},
numpages = {12},
location = {Pittsburgh, Pennsylvania},
series = {ICSE '22}
}

\end{document}